\documentclass[aps,prd,showpacs]{revtex4}

\usepackage{amsmath}
\usepackage[dvips]{graphicx}

\begin{document}

\title{%
Majorana neutrino textures from numerical considerations: \\
the CP conserving case 
\\}
\author{B. Dziewit, K. Kajda, J. Gluza, M. Zra\l ek}
\affiliation{Institute  of Physics, University of
Silesia, \\
Uniwersytecka 4, PL-40-007 Katowice, Poland }
\date{\today}

\begin{abstract}
Phenomenological bounds on the neutrino mixing matrix $U$ are used to 
determine numerically the allowed range of real elements (CP conserving case)
for the symmetric neutrino mass matrix $M_{\nu}$ 
(Majorana case). For this purpose an adaptive Monte Carlo generator has been
used. Histograms are constructed to show which forms of the
neutrino mass matrix $M_\nu$  are possible and preferred.
We confirm results found in the literature which are based on analytical 
calculations, though a few differences appear. 
These cases correspond to some textures with two zeros. The results 
show that actually both normal and
inverted mass hierarchies are still possible at $3 \sigma$ confidence level. 
\end{abstract}
\maketitle

\section{Introduction}
The standard neutrino theory involves diagonalization of the neutrino mass matrix $M_{\nu}$  by use of the  mixing matrix $U$: 

\begin{equation}
m_{diag} = U^T M_{\nu} U. \label{eq:diagonalisation}
\end{equation}

The matrices $m_{diag}$ and $U$ are restricted by experimental data. Merging information from neutrino oscillation experiments with neutrinoless double beta and
tritium decay data \cite{Czakon:1999cd}-\cite{Czakon:2001uh}, 
the absolute neutrino masses can be found to be in the range:

\begin{equation}
m_i\leq \;2.2\;{\rm eV},\;\;\;\;  
|m_i-m_j| < 0.05\;{\rm eV},\;\;\;i,j=1,2,3.
\end{equation}

while $U$ is restricted by the global neutrino oscillation analysis
\cite{Gonzalez-Garcia:2004jd}, see Table~\ref{udata}.

\begin{table}[!ht]
\begin{center}
	\begin{tabular}{c|c|c|c}
	\hline \hline
	$ i$ & $  x_{i} $ & $ x_{i}^{cent} $ & $ \sigma_{i} $\\
	\hline
	$ 1$ & $\qquad \Delta m_{32}^{2} \qquad$ & $\qquad 2.6\cdot 10^{-3} \qquad$ & $\qquad 10^{-3} \qquad$ \\
	$ 2$ & $ \Delta m_{21}^{2} $ & $ 8.3\cdot 10^{-5} $ & $ 10^{-5}$ \\
	\hline
	$ 3$ & $ |U_{e1}| $ & $ 0.835 $ & $ 0.045 $\\
	$ 4$ & $ |U_{e2}| $ & $ 0.54 $ &  $ 0.07 $\\
	$ 5$ & $ |U_{e3}| $ & $ 0.1 $ & $ 0.1 $\\
	\hline
	$ 6$ & $ |U_{\mu 1}| $ & $0.355$ & $ 0.165 $\\
	$ 7$ & $ |U_{\mu 2}| $ & $0.575$ &  $ 0.155 $\\
	$ 8$ & $ |U_{\mu 3}| $ & $0.7$ & $ 0.12 $\\
	\hline
	$ 9$ & $ |U_{\tau 1}| $ & $0.365 $ & $ 0.165 $\\
	$10$ & $ |U_{\tau 2}| $ & $0.59 $ &  $ 0.15 $\\
	$11$ & $ |U_{\tau 3}| $ & $0.685 $ & $ 0.125 $\\
	\hline
	\end{tabular}
	\caption{The allowed absolute values of the neutrino mass squared differences $\Delta m_{32}^{2}$, $\Delta m_{21}^{2}$ and the
allowed absolute values of the neutrino mixing matrix elements $|U_{ij}|$.
$x_{i}^{cent}$ and $\sigma_{i}$ are the central values and the $3 \sigma $ uncertainties, respectively.}
\label{udata}
\end{center}
\end{table}

However, what is a possible structure of the basic neutrino mass matrix $M_{\nu}$? Its knowledge could definitely shed a light 
on the mechanism of neutrino mass generation and is a subject of intensive studies \cite{Altarelli:2004za}-\cite{Merle:2006du}. 
We concentrate on theories where quark and lepton mixings are disconnected. Then a base with diagonal charged lepton mass matrix can be used and information on  neutrino mixings 
is connected solely with neutrino mass matrix $M_\nu$.
There are many papers devoted  to the problem of determination of $M_{\nu}$. Basically, they can be divided into two groups
where different methods to solve the problem are used. 
The first, ``top-down'' method relies on theoretical considerations of possible texture zeros and global symmetries
which seems to arise from the neutrino mass matrix structure, e.g. \cite{Grimus:2004hf}, \cite{Grimus:2004az},\cite{Grimus:2005sm}. 
There are also papers where the relation Eq.~\ref{eq:diagonalisation} is used to determine $M_{\nu}$ analytically 
in terms of neutrino masses and values of neutrino mixing matrix elements, e.g.  \cite{Frampton:2002yf},\cite{Desai:2002sz},\cite{Xing:2002ap},\cite{Xing:2002ta},\cite{Guo:2002ei}. 
The second, ``bottom-up'' method relies on numerical analysis. One can perform numerical diagonalization of many neutrino mass matrix textures 
leaving only these which are in agreement with present experimental data, e.g. \cite{Hall:1999sn}, \cite{Haba:2000be}, \cite{Merle:2006du}.
In this work we follow the second way: numerical solutions of neutrino mass textures are performed which are based 
on Monte Carlo simulations \footnote{A similar approach has been used in a different context to restrict nonstandard
parameters of the left-right symmetric model in \cite{Kiers:2002cz}.}. 

We have scattered a huge number  of possible texture zeros, 
our analysis confirms most of the results based on analytic approach:  
we did not find numerical solutions for textures with number of  zeros 
$n \ge 3$, we have found seven two zero textures which are in the range 
of current experimental results. Other eight two zero textures gave no results.
For some discrepancies, see the next section. 

In this paper the real symmetric $3 \times 3$ neutrino mass matrix 
is analyzed. It means that 
we assume directly the Majorana nature of neutrinos and that the investigation is restricted to the CP conserving case.
Certainly, it may seem to be a simple case, however, we believe that our visual presentation of results and
conclusions show that even this case is nontrivial. The case with CP phases will be analyzed in a separate work.
 
A general neutrino mass matrix $M_{\nu}$ which we analyze has the following form:
\begin{equation}
	\left(
	\begin{array}{ccc}
	a & b & c\\
	b & d & e\\
	c & e & f
	\end{array}
	\right). \label{eq:generalcase}
\end{equation}
It has six parameters.
Our approach is the following. First we consider the general case where all six parameters in Eq.~\ref{eq:generalcase}
can be nonzero. Taking into account data in Table~\ref{udata},  we found the allowed range of these parameters.  
Histograms (frequency spectrums) for neutrino mass matrix elements (both for normal and inverted mass hierarchies)
have been also constructed. These histograms give a nice visualization of possible solutions for $M_{\nu}$ and give a hint which regions 
of parameters are in favor (e.g. for texture zeros). Next, a detailed analysis of the $M_{\nu}$ textures are made where some of the parameters are exactly zero. Finally, conclusions are given  and the Appendix contents details on the Monte Carlo algorithm used in the paper.

\section{The general case}
Let us perform first numerical fitting for the general case of the $M_{\nu}$ neutrino mass matrix (all parameters non-zero). 
To find numerical solutions, the matrix Eq.~\ref{eq:generalcase} with some randomly generated numerical elements is diagonalized 
and the eigensystem is found: 
eigenvalues (neutrino masses) and normalized eigenvectors (columns of the neutrino mixing matrix $U$). 
Using eigenvalues,  neutrino mass squared differences can be constructed: $\Delta m_{21}^{2}$ and $\Delta m_{32}^{2}$. 
Both $\Delta m_{21}^{2}$, $\Delta m_{32}^{2}$ and elements of $U$ (let us call them $x_i$ points) are compared with experimental values written in Table~\ref{udata} ($x_i^{cent}$). 
The rejection rule for a considered texture is defined in a form:
\begin{equation}
	\frac{(x_{i}^{cent}-x_{i})^2}{\sigma_{i}^{2}} \le \frac{1}{\alpha^{2}}.
\label{eq:ch}
\end{equation} 
If inequality Eq.~\ref{eq:ch} is fulfilled, then the parameters $x_i$ are possible solutions of the problem. 
To find the best values of parameters, the whole procedure is more refined, we send a reader to the Appendix A where 
the whole algorithm based on Adaptive Monte Carlo method is described.

The parameter $\alpha$ gives an information about decreasing of the assumed error $\sigma_{i}$. 
If $\alpha = 1$, then experimental data from the Table~\ref{udata} are taken and results are obtained at $3 \sigma$ confidence level.
 If $\alpha > 1$ then we assume that errors shrink, however, 
we still assume the central values $x_i^{cent}$ of Table~\ref{udata}. In this way $\alpha$ informs us how much $x_i$ solutions are focused around 
their central values (see section III) . 

Let us note that modulus of the neutrino mixing matrix elements $|U_{ij}|$ are analyzed and it is not possible to distinguish 
rotation symmetries of angles $\theta_{ij}$, as a consequence, the figures included here have symmetric regions of allowed parameters.

The algorithm is able to distinguish the normal mass hierarchy from the inverted one 
by finding only such combinations of neutrino mass matrix parameters (textures) which imply neutrino masses  
$m_{1}<m_{2}\ll m_{3}$ (normal hierarchy) and $m_{3}\ll m_{1}<m_{2}$ (inverted hierarchy).
In this way thousands of sets of allowed parameters can be find for a given hierarchy, 
these are plotted in Fig.~\ref{gen1}. To allow for easier visualization of results, three dimensional projections of regions on the basic planes are  
shown (gray colours). 
In Fig.~\ref{gen1f} histograms for all $M_\nu$ mass matrix parameters are shown. 
One can notice that histograms for normal mass hierarchy with $a,b,c$ parameters have their maxima around zero.
The opposite situation takes place for inverse mass hierarchy Fig.~\ref{gen2f}. 
Here the zero elements are connected with  $d$, $e$ and $f$ parameters.
 These solutions suggest that  textures with some parameters set to zero should be analyzed with a greater care. 
It will be done in the next section.

\section{Texture zeros}
Here we will assume some parameters of the matrix $M_{\nu}$ to be zero, exactly. It appears, that taking into account present 
experimental data Table~\ref{udata} 
we got in most cases solutions with both normal and inverse mass hierarchies.
For an interesting case when $a=0$ (it means that the effective neutrinoless double beta decay mass parameter $<m_{\beta \beta 0 \nu}>$, 
see e.g. \cite{Czakon:1999cd}-\cite{Czakon:2001uh} is zero for this texture) only normal mass hierarchy is possible (texture A). Results are given in Fig.~\ref{azero}. 
In this figure, the second row gives solutions when experimental errors are
shrinking two times, without changing central values $x_i^{cent}$. We can see, that regions decreased drastically. It means that many of 
the solutions found are placed nearby their central values and many of peripheral solutions connected with edged errors are cut 
away by our rejection rule given by the $\alpha$ parameter.
In Fig.~\ref{azeroH} the solutions are transformed to the physical parameters 
connected with mixing angles. 

Coming from the  numerical solutions of the mixing matrix $U$
into the language of mixing angles, the standard form of the
neutrino mixing matrix represented
by three angles $\theta_{13}$, $\theta_{23}$ and $\theta_{12}$
 has been used (see e.g. \cite{Gonzalez-Garcia:2004jd}, CP phase $\delta=0$).
If a specific hierarchy of the neutrino masses is chosen,
the mixing angles defined by oscillation probabilities 
can be restricted to the first quadrant
\cite{Gluza:2001de}. However, in Fig.~\ref{azeroH} (and also Figs. 
\ref{CtexH}-\ref{BH}) 
we consider general approach with unbounded  mixing angles.
We can see that some of histograms are not symmetric against
$\sin{\theta_{ij}}=0$. 
In Fig. 6 we keep the plot of $\sin{\theta_{23}}$, though for the
experimentally and phenomenologically interesting case $a=0$ the
neutrinoless double beta decay is independent of $\theta_{23}$ 
(see e.g. \cite{Xing:2003jf,Lindner:2005kr} ).
It means that the same histogram results for the case
when $a \neq 0$.
Values of $\sin{\theta_{ij}}$ parameters in 
Figs.~\ref{azeroH},\ref{CtexH}-\ref{BH} 
correspond to the 3 $\sigma$ confidence level.
We can see in Fig.~\ref{azeroH} that $\theta_{13}$ around
zero is preferred for this texture.
Histograms for all other cases with a one zero parameter can be found in addition in \cite{www}. 
There are analytical considerations for the case $a=0$ in the literature, 
see e.g. \cite{Xing:2003jf,Lindner:2005kr}.
It is easy to prove  that in the case $a=0$, and for $\theta_{13}=0$ 
a simple relation holds
\begin{equation}
\left| \frac{m_1}{m_2} \right| = \tan^2 {\theta_{12}}.
\end{equation}
We have checked that our numerical solutions fulfill this relation.

Let us discuss results for the  case where two  zero parameters are presented in Eq.~\ref{eq:generalcase}.

If $n$ of six independent parameters in Eq.~\ref{eq:generalcase} vanish then one can find $\frac{6!}{n!(6-n)!}$ possible textures. For $n=2$, we have 15 cases.
However, we have found that, to be in agreement with present experimental data (i.e. $\alpha=1$, $3 \sigma$ c.l.), 
seven  cases are possible.
All possible two zero textures are shown in Table~\ref{2t}. 
In the case of the normal mass hierarchy, we have got the so called 
$A_{1}$ and $A_{2}$ textures \footnote{The names taken after \cite{Frampton:2002yf}, \cite{Desai:2002sz}, \cite{Xing:2004ik} and \cite{Lavoura:2004tu}} with $a,b=0$ and $a,c=0$, respectively.
$B$ textures have both normal and inverse hierarchies and C texture has inverted one (see Table~\ref{2zero}). 
From the general case (inverted mass hierarchy in the last section) we have concluded that $d,e,f$ parameters can be zero, 
however, it appears that only situation when $d=0$ and $f=0$ is possible, all other combinations, e.g. $d=0$ and $e=0$ are excluded
(to come to this conclusion, altogether around $10^9$ randomly generated points have been used for each degree of freedom).  
In Table~\ref{2zero} we recalculated solutions for $M_\nu$ to get physical values of neutrino masses.
The last column shows results for the $\alpha_0$ parameter for which no solutions have been found.
As already discussed, smaller $\alpha_0$, more parameters around zero are allowed.
The calculation is stopped when $10^9$ points for each
degree of freedom for the Monte Carlo random generator is not able to find solutions in agreement with Eq.~\ref{eq:ch}.
The one before last column in the Table~\ref{2zero} shows a prediction for the neutrinoless double beta decay parameter $<m_{\beta \beta 0 \nu}>$. We will consider 
this important parameter in more details in a separate paper, where also possible CP violating phases will be examined.

\begin{table}[!ht]
\begin{center}
	\begin{tabular}{c|c}
	\hline \hline
	$ \qquad TEXTURE \qquad $ & $ \qquad ZERO \quad PARAMETERS \qquad $ \\
	\hline
	$ A_{1} $ & $ a,b=0 $  \\
	$ A_{2} $ & $ a,c=0 $  \\
	\hline
	$ B_{1} $ & $ c,d=0 $ \\
	$ B_{2} $ & $ b,f=0 $  \\
	$ B_{3} $ & $ b,d=0 $ \\
	$ B_{4} $ & $ c,f=0 $  \\
	\hline
	$ C $ & $ d,f=0 $  \\
	\hline
	\end{tabular}
	\caption{The table shows all two zero textures which are in agreement with current experimental data.}
\label{2t}
\end{center}
\end{table}

\begin{table}[!ht]
\begin{center}
	\begin{tabular}{c|c|c|c|c}
	\hline \hline
	$ TEXTURE $ &  $ ZERO \quad PARAMETERS $ & $ MASS \quad RANGE $ & $ MASS\quad MEAN $ & $ \alpha_{0} $\\
	\hline
	$ A     $ & $ a  =0 $ &  $\mid m_{3}\mid=(0.041,0.062)$ &$<m_{3}>=0.052$ & 3.32\\
	normal    &           &  $\mid m_{2}\mid=(0.009,0.015)$ &$<m_{2}>=0.010$ & \\
	          &           &  $\mid m_{1}\mid=(0.002,0.011)$ &$<m_{1}>=0.005$ & \\
	          &           &  $\mid m_{\beta \beta 0 \nu}\mid=0$ &$<m_{\beta \beta 0 \nu}>=0$ & \\
	\hline
	$A_{1}$ and $A_2$ & $ a,b=0 $ &  $\mid m_{3}\mid=(0.041,0.062)$ &$<m_{3}>=0.053$ & 2.65\\
	normal    & $ a,c=0 $ &  $\mid m_{2}\mid=(0.009,0.015)$ &$<m_{2}>=0.011$ & \\
	          &           &  $\mid m_{1}\mid=(0.002,0.012)$ &$<m_{1}>=0.004$ & \\
	          &           &  $\mid m_{\beta \beta 0 \nu}\mid=0$ &$<m_{\beta \beta 0 \nu}>=0$ & \\
	\hline 
  $ B_{1},B_{2} $ & $ c,d=0 $ &  $\mid m_{3}\mid=(0.05,0.14)$ &$<m_{3}>=0.08$ & 1.18\\
       degenerate & $ b,f=0 $ &  $\mid m_{2}\mid=(0.03,0.13)$ &$<m_{2}>=0.06$ & \\
	   or     &           &  $\mid m_{1}\mid=(0.02,0.13)$ &$<m_{1}>=0.06$ & \\
	 normal   &           &  $\mid m_{\beta \beta 0 \nu}\mid=(0.02,0.13)$ &$<m_{\beta \beta 0 \nu}>=0.06$ & \\
	\hline
  $ B_{1},B_{2} $ & $ c,d=0 $ &  $\mid m_{2}\mid=(0.05,0.18)$ &$<m_{2}>=0.09$ & 1.18 \\
       degenerate & $ b,f=0 $ &  $\mid m_{1}\mid=(0.05,0.18)$ &$<m_{1}>=0.09$ & \\
	   or     &           &  $\mid m_{3}\mid=(0.03,0.17)$ &$<m_{3}>=0.07$ & \\
	inverted  &           &  $\mid m_{\beta \beta 0 \nu}\mid=(0.05,0.18)$ &$<m_{\beta \beta 0 \nu}>=0.09$ & \\
	 \hline 
  $ B_{3},B_{4} $ & $ b,d=0 $ &  $\mid m_{3}\mid=(0.05,0.22)$ &$<m_{3}>=0.08$ & 1.25\\
       degenerate & $ c,f=0 $ &  $\mid m_{2}\mid=(0.025,0.21)$ &$<m_{2}>=0.06$ & \\
	   or     &           &  $\mid m_{1}\mid=(0.02,0.205)$ &$<m_{1}>=0.06$ & \\
	normal    &           &  $\mid m_{\beta \beta 0 \nu}\mid=(0.03,0.21)$ &$<m_{\beta \beta 0 \nu}>=0.06$ & \\
	\hline
  $ B_{3},B_{4} $ & $ b,d=0 $ &  $\mid m_{2}\mid=(0.05,0.25)$ &$<m_{2}>=0.083$ & 1.25\\
       degenerate & $ c,f=0 $ &  $\mid m_{1}\mid=(0.045,0.25)$ &$<m_{1}>=0.082$ & \\
	   or     &           &  $\mid m_{3}\mid=(0.03,0.24)$ &$<m_{3}>=0.065$ & \\
	inverted  &           &  $\mid m_{\beta \beta 0 \nu}\mid=(0.045,0.246)$ &$<m_{\beta \beta 0 \nu}>=0.084$ & \\
	\hline 
	$ C $     & $ d,f=0 $ &  $\mid m_{2}\mid=(0.042,0.072)$ &$<m_{2}>=0.056$ & 2.65\\
	inverted  &           &  $\mid m_{1}\mid=(0.041,0.071)$ &$<m_{1}>=0.055$ & \\
	          &           &  $\mid m_{3} \mid=(0.012,0.039)$ &$<m_{3}>=0.023$ & \\
	          &           &  $\mid m_{\beta \beta 0 \nu}\mid=(0.011,0.039)$ &$<m_{\beta \beta 0 \nu}>=0.022$ & \\
	\hline
	\end{tabular}
	\caption{Masses and effective neutrinoless double beta decay mass parameter $<m_{\beta \beta 0 \nu}>$ 
for allowed textures with two zeroes and one with one zero.
The last column shows the parameter $\alpha_0$ for which schemes have no positive solutions.}
\label{2zero}
\end{center}
\end{table}

\begin{table}[!ht]
\begin{center}
	\begin{tabular}{c|c|c|c|c}
	\hline \hline
	$ i$ & $  \qquad x_{i} \qquad $ & $ \qquad A_1,A_2 \quad $ & $ \qquad B_1-B_4 \qquad $ & $ \qquad C \qquad    $ \\
	\hline
	$ 1$ & $  \Delta m_{32}^{2}   $ & $       2.65 \cdot 10^{-3}   $ & $        2.55 \cdot 10^{-3} $ & $ 2.61 \cdot 10^{-3} $ \\
	$ 2$ & $  \Delta m_{21}^{2}   $ & $       8.27 \cdot 10^{-5}   $ & $        8.35 \cdot 10^{-5} $ & $ 8.30 \cdot 10^{-5} $ \\
	\hline
	$ 3$ & $ |U_{e1}|             $ & $       0.84                 $ & $        0.84               $ & $ 0.84               $ \\
	$ 4$ & $ |U_{e2}|             $ & $       0.54                 $ & $        0.54               $ & $ 0.54               $ \\
	$ 5$ & $ |U_{e3}|             $ & $       0.12                 $ & $        1.9 \cdot 10^{-3}  $ & $ 0.06               $ \\
	\hline
	$ 6$ & $ |U_{\mu 1}|          $ & $       0.41                 $ & $        0.40               $ & $ 0.36               $ \\
	$ 7$ & $ |U_{\mu 2}|          $ & $       0.56                 $ & $        0.63               $ & $ 0.58               $ \\
	$ 8$ & $ |U_{\mu 3}|          $ & $       0.72                 $ & $        0.66               $ & $ 0.72               $ \\
	\hline
	$ 9$ & $ |U_{\tau 1}|         $ & $       0.36                 $ & $        0.36               $ & $ 0.41               $ \\
	$10$ & $ |U_{\tau 2}|         $ & $       0.63                 $ & $        0.55               $ & $ 0.59               $ \\
	$11$ & $ |U_{\tau 3}|         $ & $       0.68                 $ & $        0.75               $ & $ 0.68               $ \\
	\hline
	\end{tabular}
	\caption{This table shows $x_{i}^{cent}$ values obtained from numerical solutions for two zero textures. It appears, that cases $A_1$ and $A_2$
coincide with solutions for one zero texture with $a=0$.}
\label{bestcent}
\end{center}
\end{table}

 From the Table~\ref{2zero} we can deduce (see the best values in the MASS MEAN column) that neutrino masses fulfill relations:

\begin{eqnarray}
m_1 &< &m_2 << m_3,\; {\rm texture\; A} \\
m_3&<<& m_1\; < \;m_2,\;\;  {\rm texture\; C},
\end{eqnarray}
in agreement with analytical considerations given in \cite{Frampton:2002yf} and \cite{Desai:2002sz}. However, for  textures B the situation 
is more complicated. In the above cited papers a degeneration of neutrino  
masses for textures B have been inferred. However, as we can see from 
Table~\ref{2zero}, for textures B not only the complete degeneration
of neutrino masses $m_1 \simeq m_2 \simeq m_3$ is allowed.
Actually, both inverted and normal mass hierarchies are possible.
We think that the reason for discrepancies lies at assumptions 
or approximations made in the analytical approach which cut off some
still possible solutions. 
Let us note, that most of analytical results to which we refer 
include cases with CP phases.
Here we have restricted ourselves to the real neutrino mass matrix cases, 
so we can expect even wider spectrum of solutions when complex 
mass matrices will be included.

In Table~\ref{bestcent} the results with the smallest $\chi^2$ 
values found by our adaptive Monte Carlo procedure 
are written. We can see that some of them ``walked away'' from 
corresponding values in Table~\ref{udata}.

Finally, in Fig.~\ref{Ctex} we visualize results for the C texture. Corresponding histograms are shown in Fig.~\ref{CtexH}.
In Fig.~\ref{AH} and Fig.~\ref{BH} the histograms for $A$ and $B$ textures are given, respectively.
We can see that for schemes $A$, nonzero $\theta_{13}$ angles are favorable.
Similarly to the case with a one zero texture, additional set of figures can be found 
in \cite{www}.

We have found, that there are no allowed textures when 3 (and more) parameters of the matrix $M_{\nu}$ are exactly zero.
This situation can be different when CP violating phases are taken into account (work in progress).

\section{Conclusions}
We have analyzed the special case of Majorana neutrino textures which give CP conserving effects.
Regions of parameters have been found which are in agreement with present experimental data. 
Allowing for random generation of parameters, histograms suggest some preferable regions of parameters.
Moreover, the results show that neutrino textures with some zero elements 
are the most probably solutions.

Let us summarize the most important points:
\begin{itemize}
  \item for the general case, 
some elements of the neutrino mass matrix $M_\nu$  
are likely to be around zero,
  \item there are no possible numerical solutions for $M_{\nu}$ textures 
with number of zeros $n \ge 3$ (at $3 \sigma$ c.l.),
  \item there are seven two zero textures which give results 
in agreement with present experimental data, some of them can have both
normal/degenerate and inverse/degenerate mass hierarchies, 
some of them have only normal, or only inverted mass hierarchies,
  \item textures $B$ give small values of $\sin \theta_{13}$,
  \item cases with $m_{\beta \beta 0 \nu}=0$  have got only normal mass 
hierarchy and they all imply similar results.
\end{itemize}

\section*{Acknowledgments}
The work  was supported in part by  the Polish State Committee for Scientific Research (KBN)
  for the research project 1P03B04926.

\appendix
\section{Description of the Adaptive Monte Carlo algorithm}
\begin{figure}[!ht]
\begin{center}
	\begin{tabular}{cc}
		\resizebox{75mm}{!}{\includegraphics{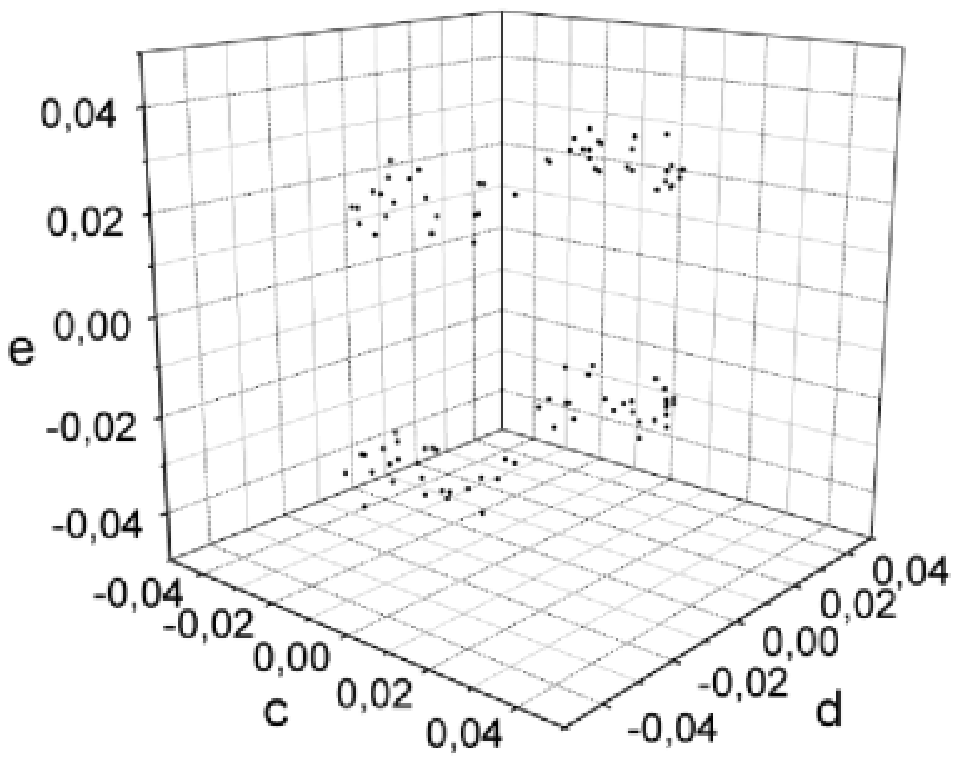}}&
		\resizebox{75mm}{!}{\includegraphics{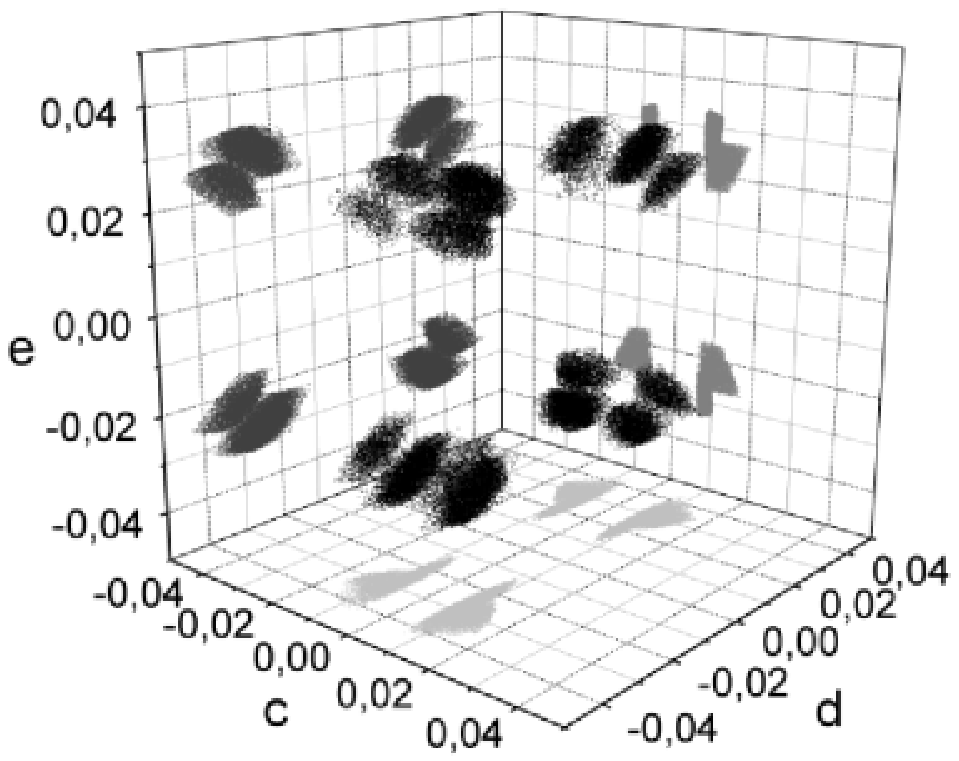}}
	\end{tabular}
	\caption{Three dimensional plots of allowed parameters found by the AMC procedure.
On the left plot there are points obtained firstly by generating random parameters which fullfill relation Eq.~\ref{eq:A1}, on the right plot the points are denser 
as AMC looks for additional solutions in a vicinity of the parameters obtained in the first step Eqs.\ref{eq:A2}-\ref{eq:A4}. These plots show solutions for the $A_{1}$ texture.}
\label{mc}
\end{center}
\end{figure}

\begin{figure}[!ht]
\begin{center}
	\begin{tabular}{cc}
		\resizebox{80mm}{!}{\includegraphics{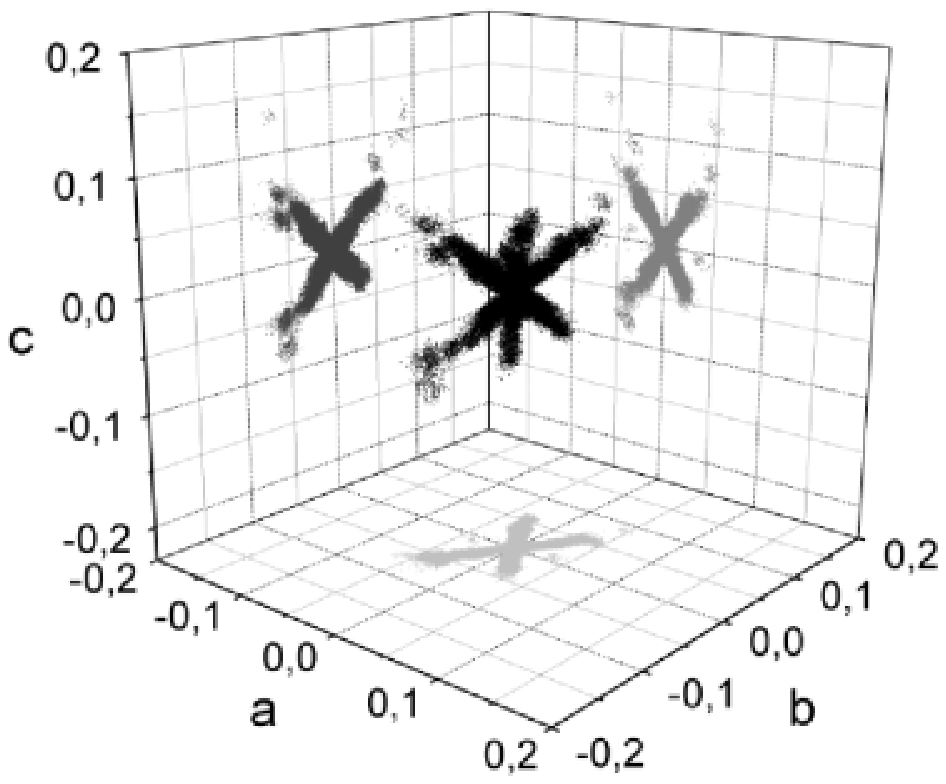}}&
		\resizebox{80mm}{!}{\includegraphics{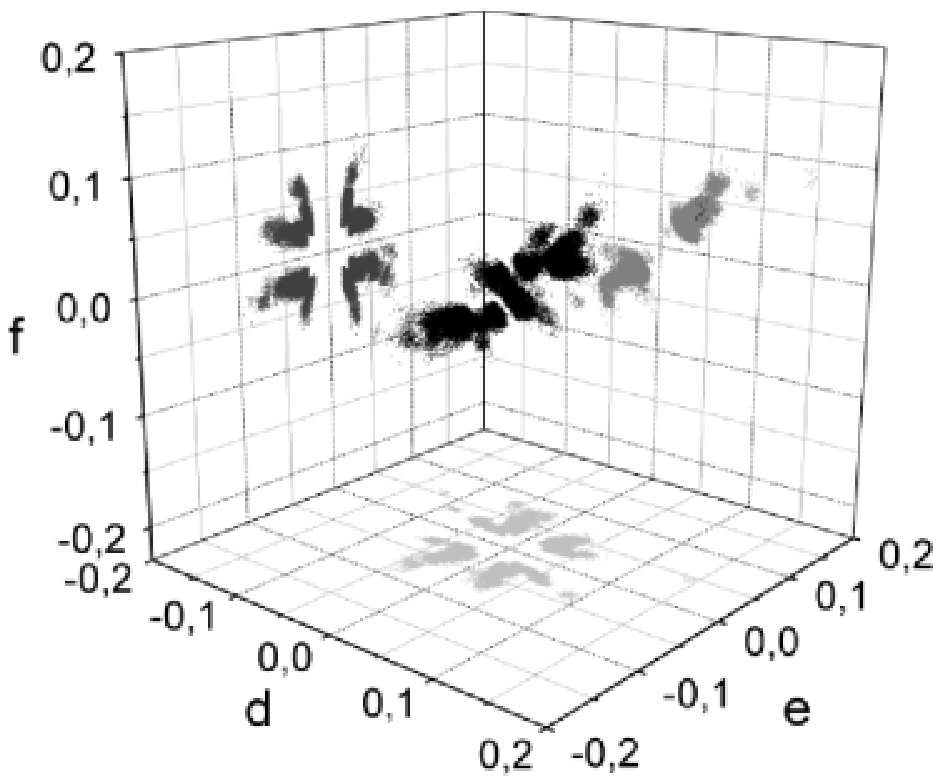}}\\
		\resizebox{80mm}{!}{\includegraphics{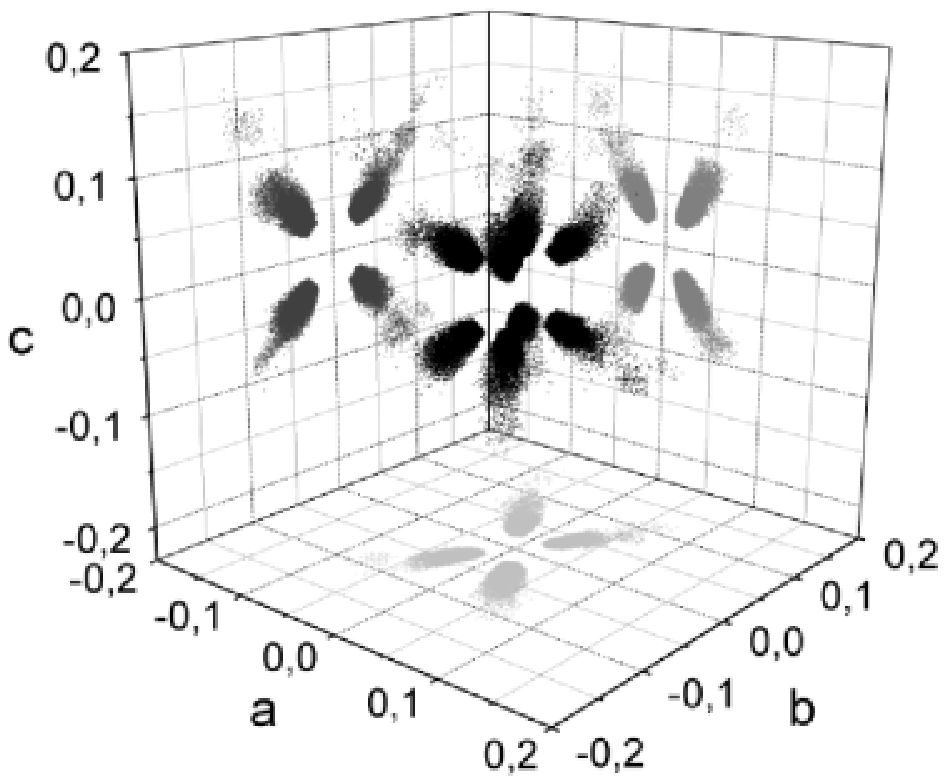}}&
		\resizebox{80mm}{!}{\includegraphics{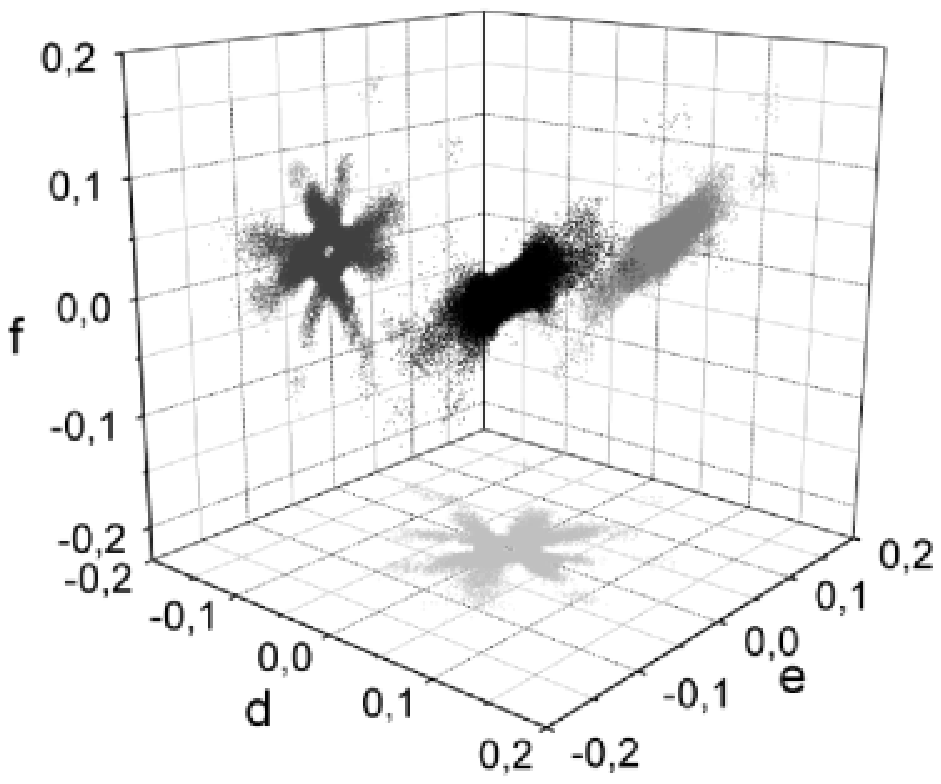}}
	\end{tabular}
	\caption{Allowed regions of parameters for the neutrino mass matrix $M_{\nu}$ with present experimental data (Table~\ref{udata}, $3 \sigma$ level), the general case with normal mass hierarchy
(upper plots) and inverted mass hierarchy (bottom plots).}
\label{gen1}
\end{center}
\end{figure}

\begin{figure}[!ht]
\begin{center}
	\begin{tabular}{ccc}
		\resizebox{60mm}{!}{\includegraphics{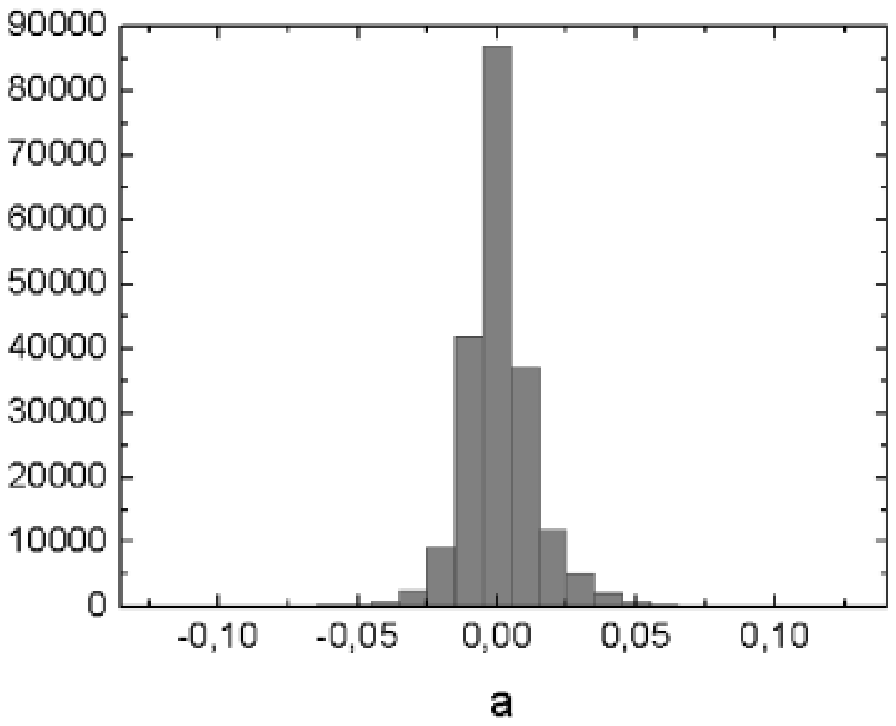}}&
		\resizebox{60mm}{!}{\includegraphics{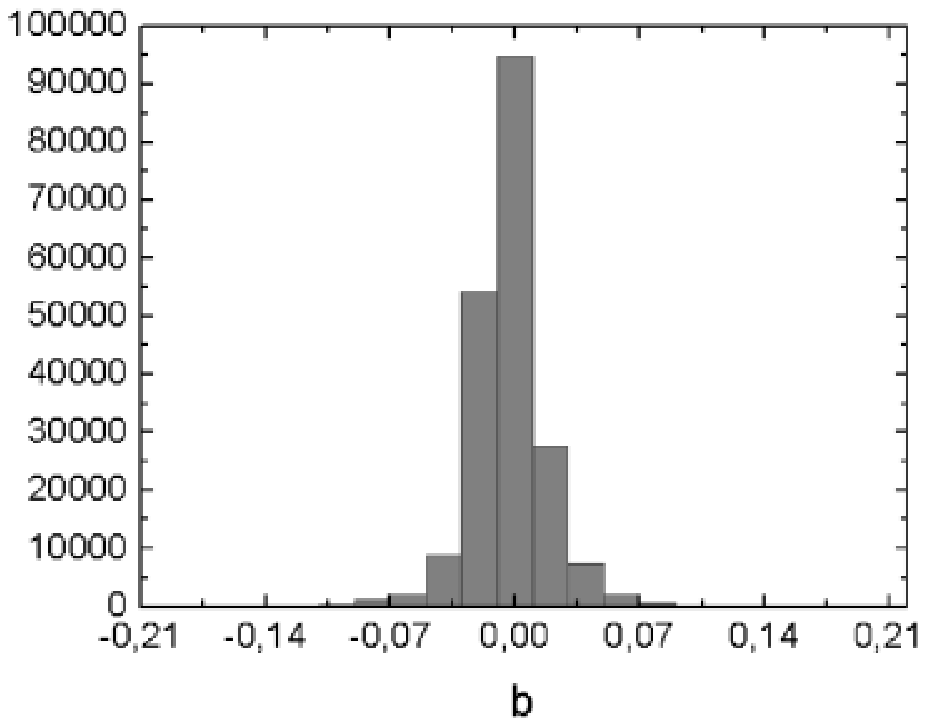}}&
		\resizebox{60mm}{!}{\includegraphics{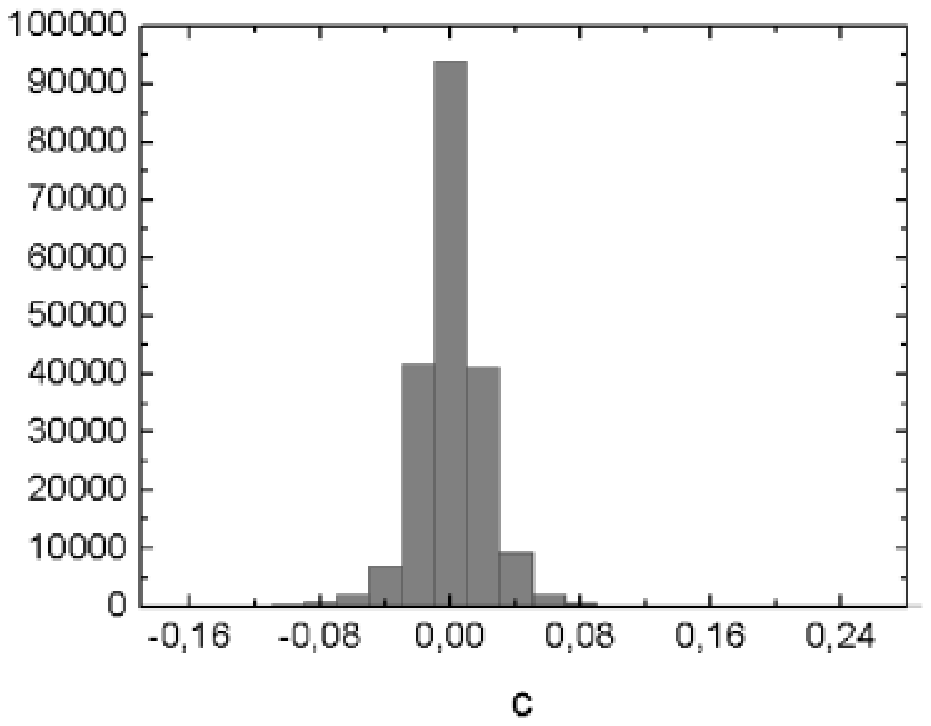}}\\
		\resizebox{60mm}{!}{\includegraphics{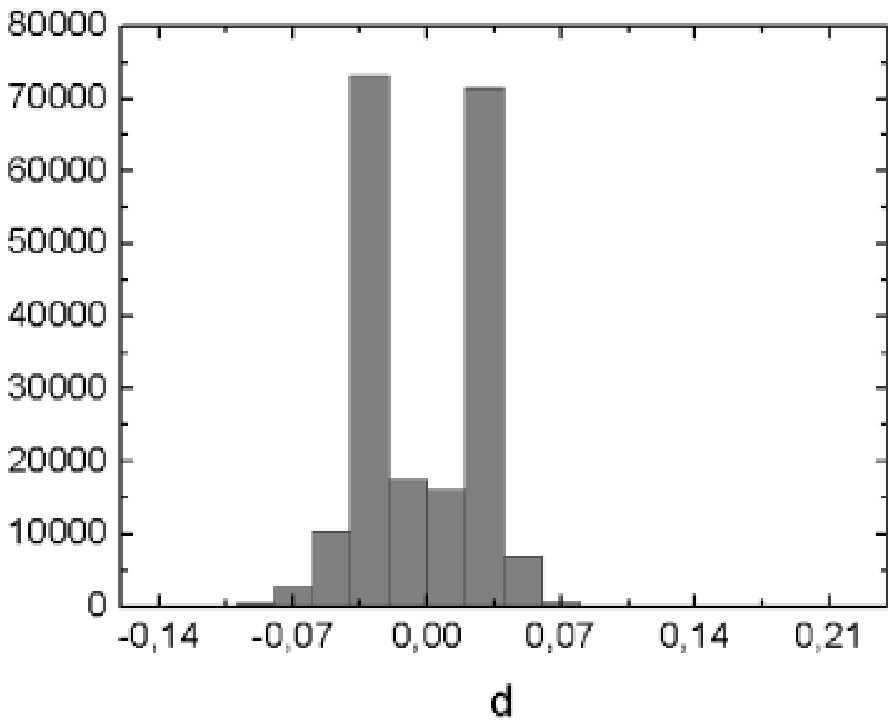}}&
		\resizebox{60mm}{!}{\includegraphics{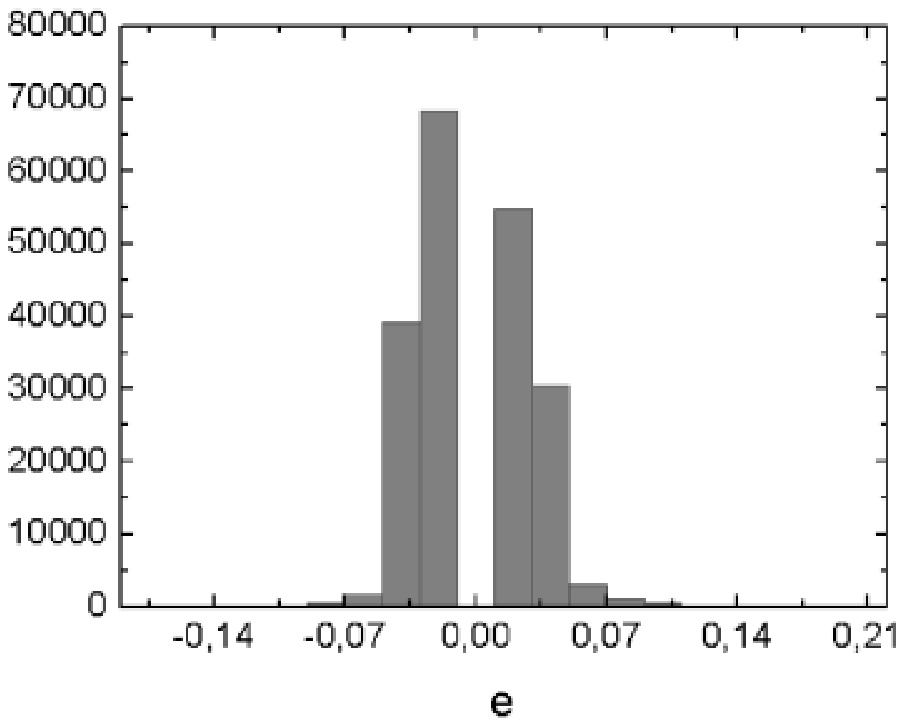}}&
		\resizebox{60mm}{!}{\includegraphics{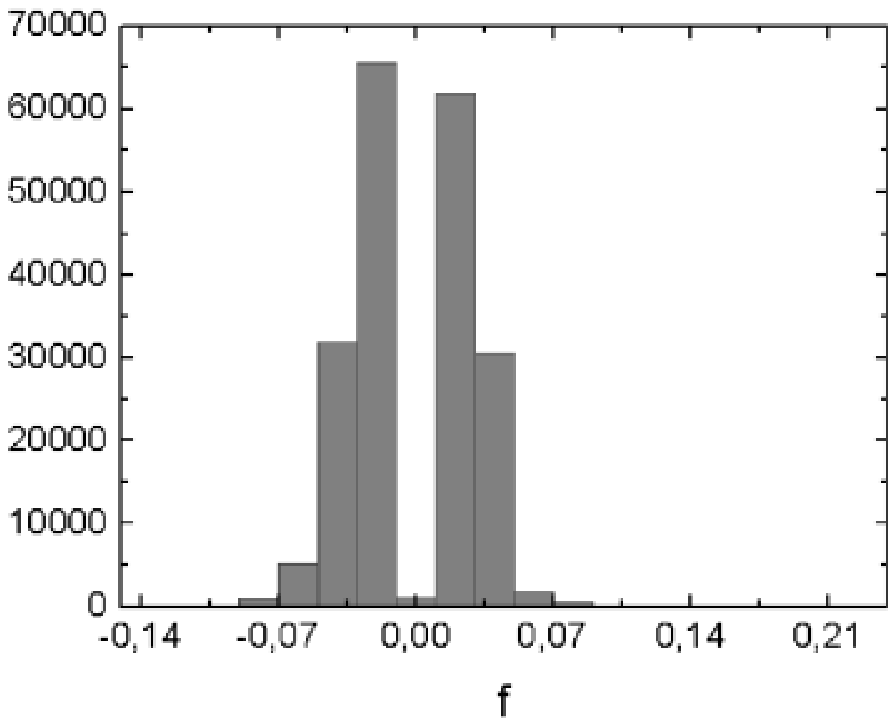}}
	\end{tabular}
	\caption{Frequency spectrum for the elements of the neutrino mass matrix $M_{\nu}$: the general case with 
normal mass hierarchy.}
\label{gen1f}
\end{center}
\end{figure}

\begin{figure}[!ht]
\begin{center}
	\begin{tabular}{ccc}
	        \resizebox{60mm}{!}{\includegraphics{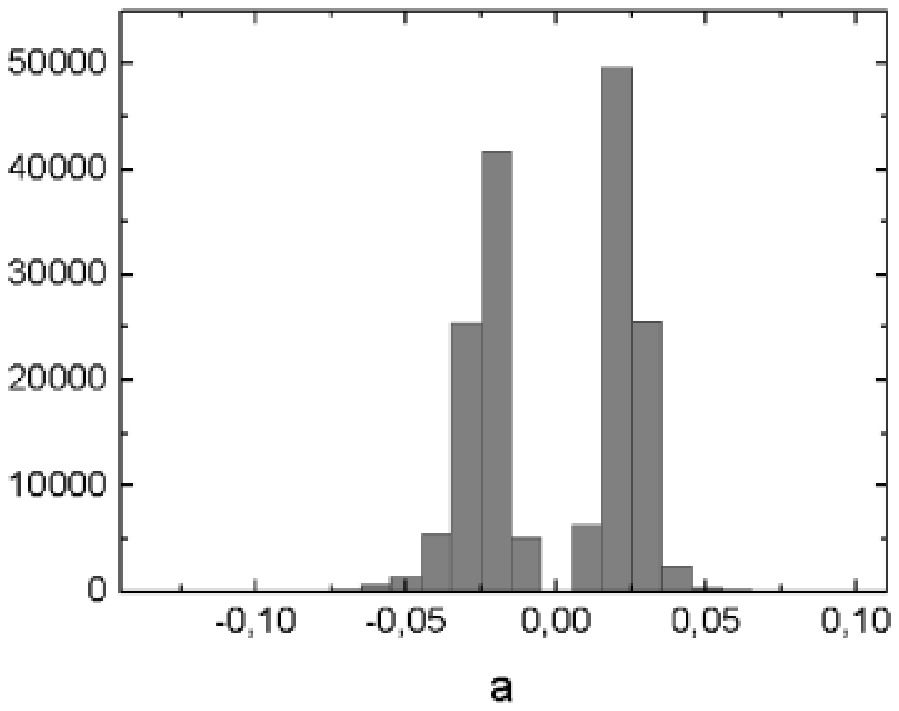}}&
		\resizebox{60mm}{!}{\includegraphics{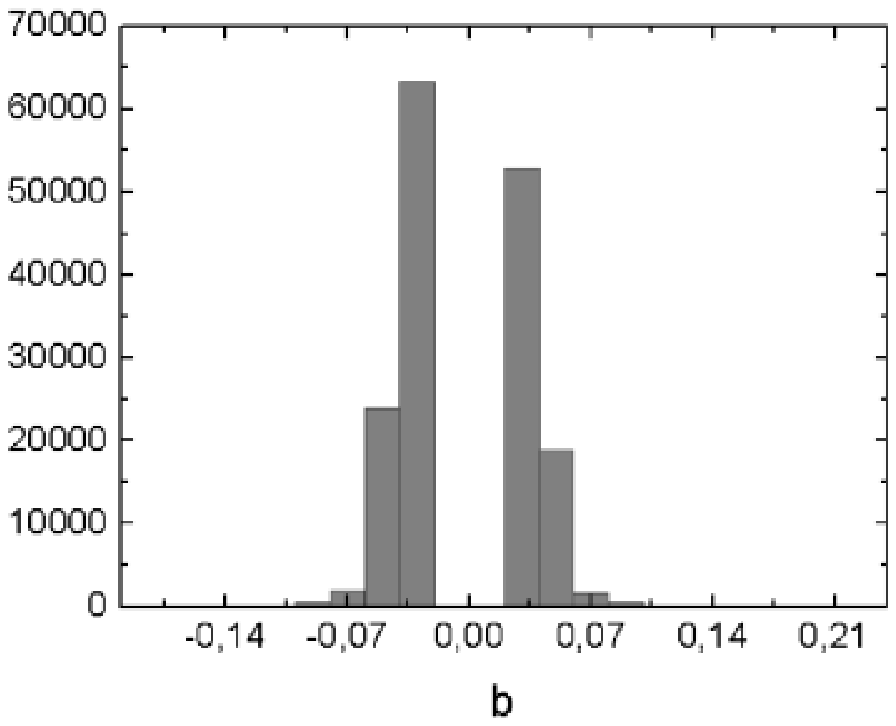}}&
		\resizebox{60mm}{!}{\includegraphics{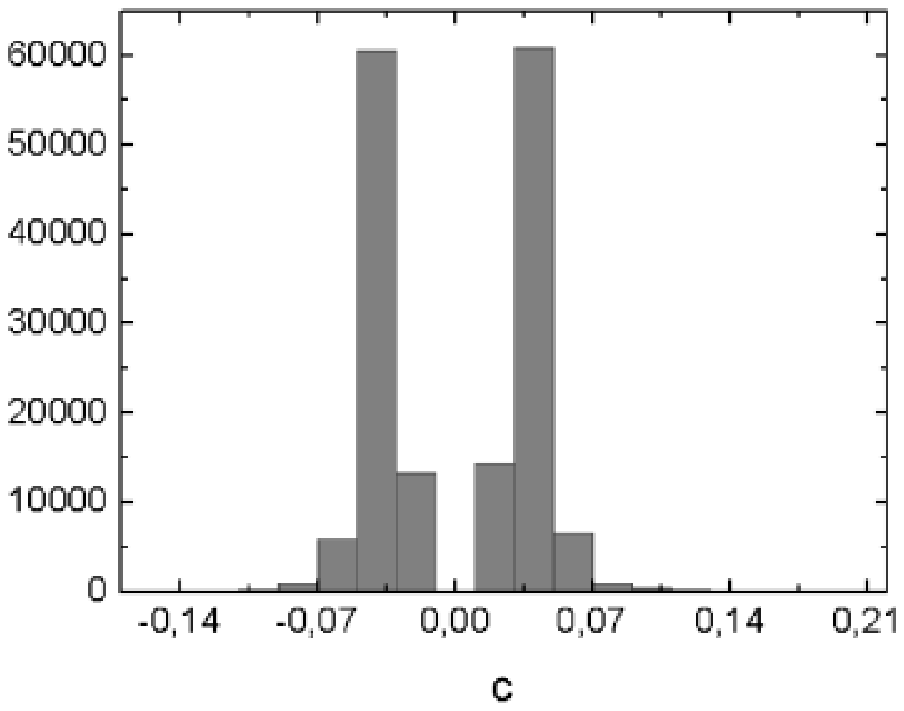}}\\
		\resizebox{60mm}{!}{\includegraphics{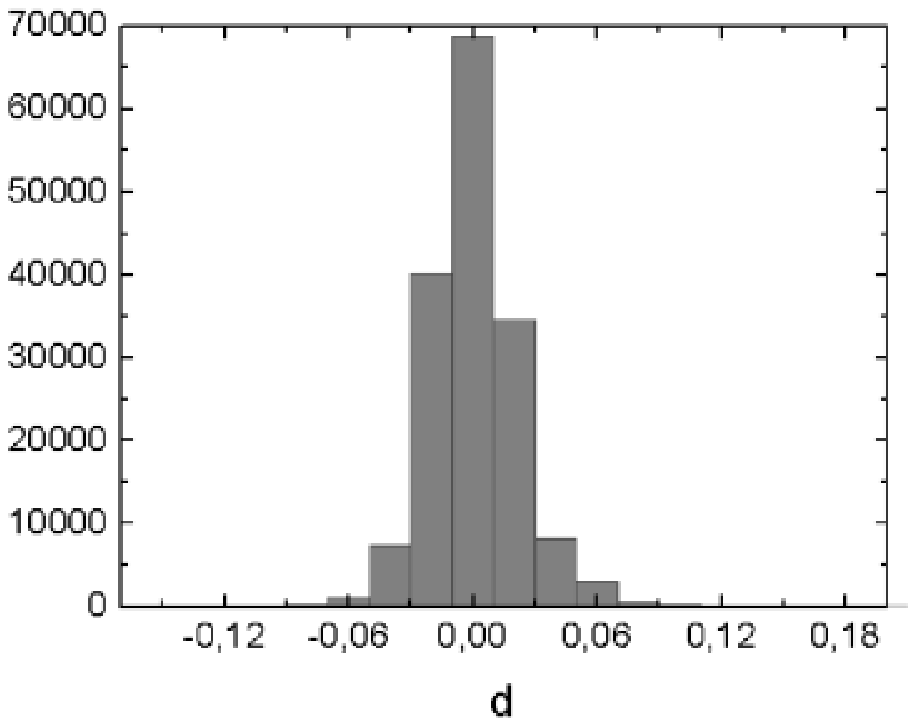}}&
		\resizebox{60mm}{!}{\includegraphics{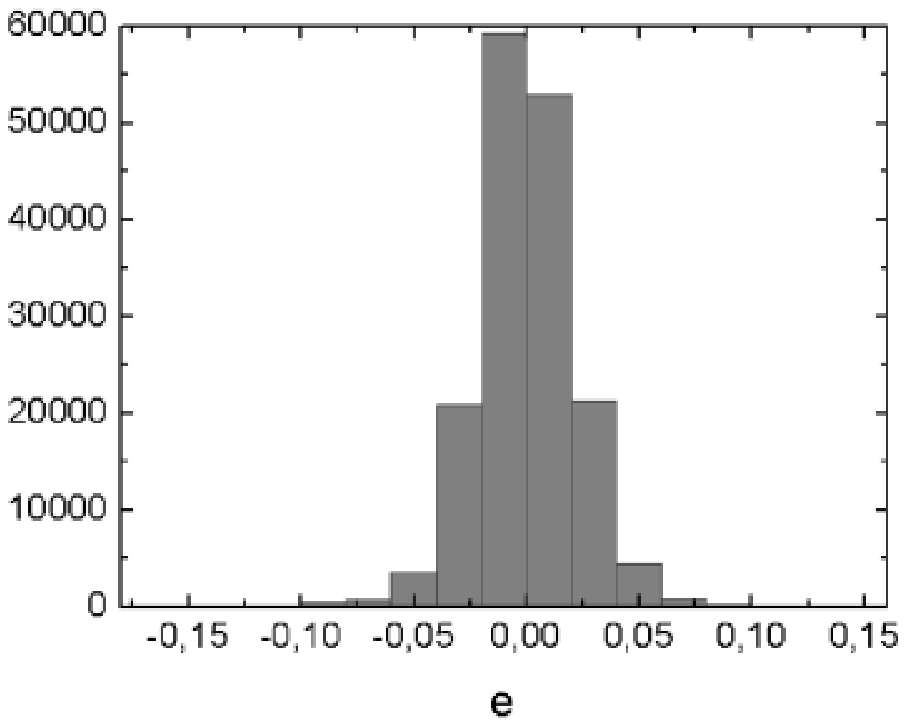}}&
		\resizebox{60mm}{!}{\includegraphics{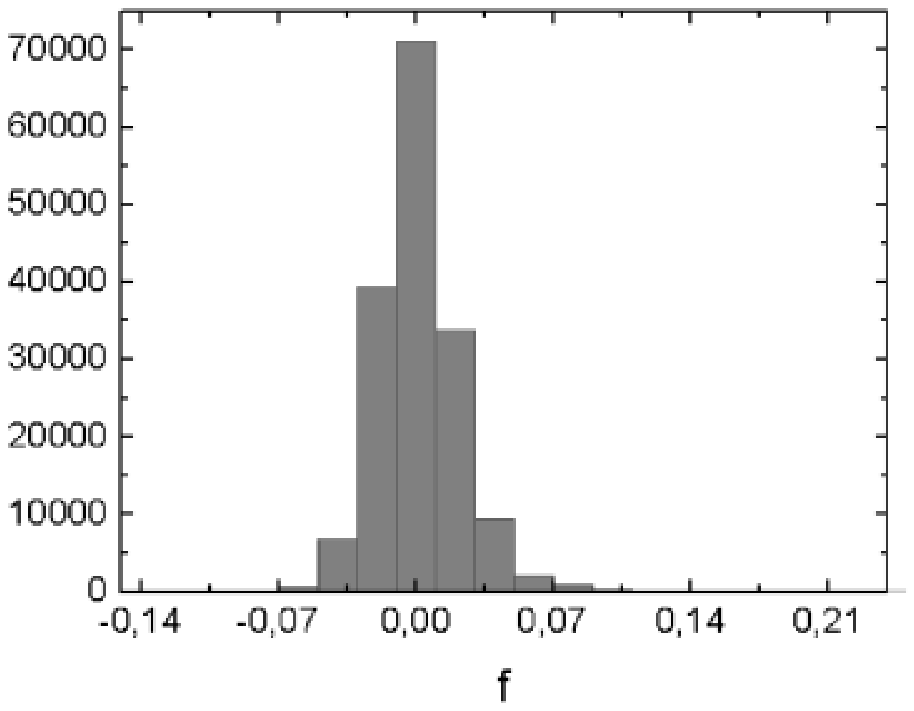}}
	\end{tabular}
	\caption{Frequency spectrum for the elements of the neutrino mass matrix $M_{\nu}$: the general case with 
inverted mass hierarchy.} 
\label{gen2f}
\end{center}
\end{figure}

\begin{figure}[!ht]
\begin{center}
	\begin{tabular}{ccc}
  	        \resizebox{80mm}{!}{\includegraphics{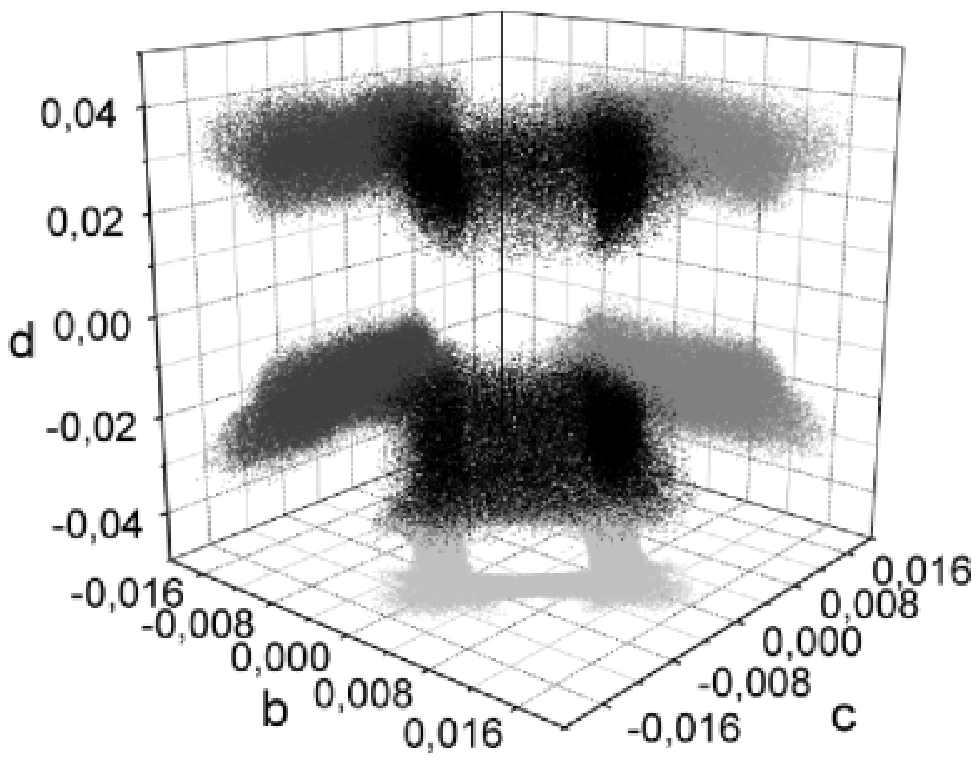}}&
		\resizebox{80mm}{!}{\includegraphics{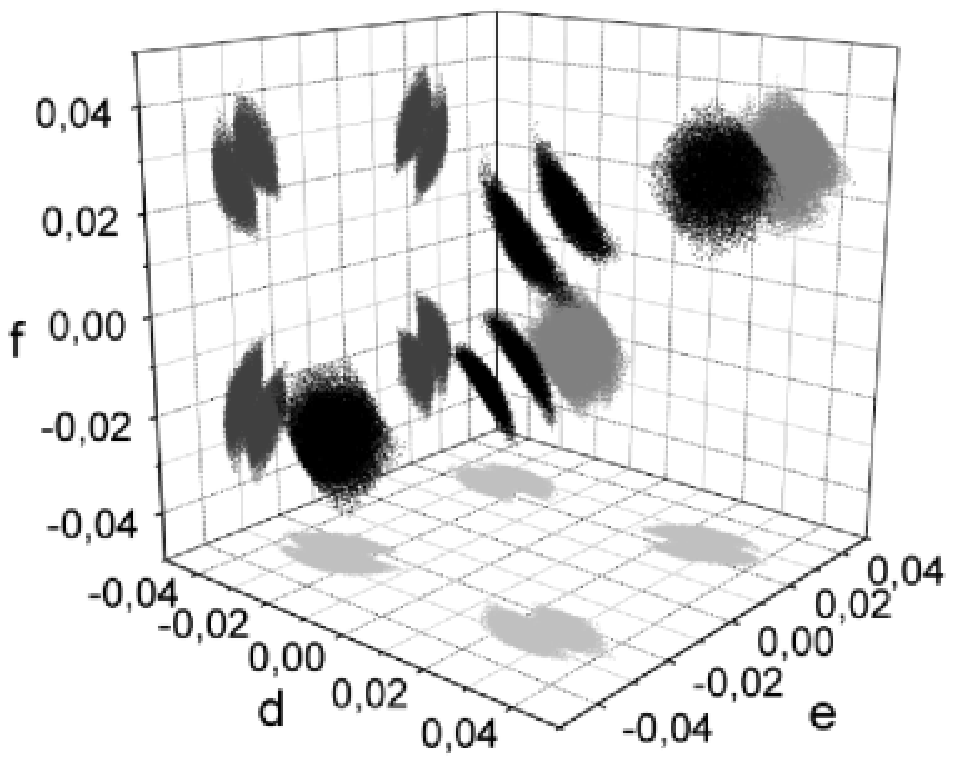}}\\
		\resizebox{80mm}{!}{\includegraphics{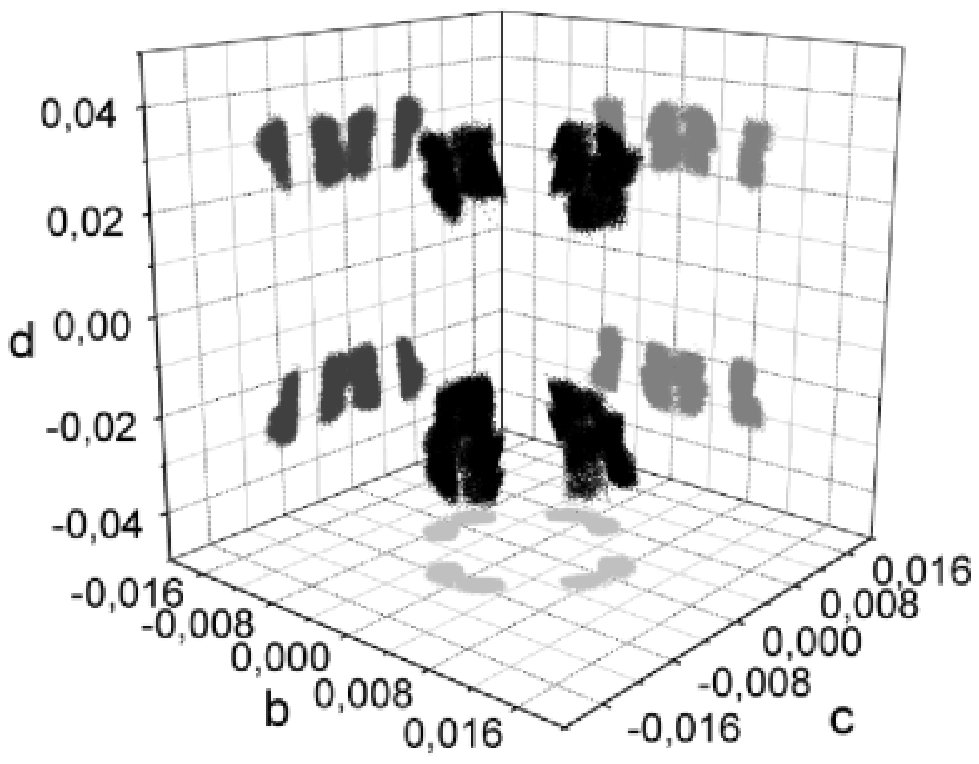}}&
		\resizebox{80mm}{!}{\includegraphics{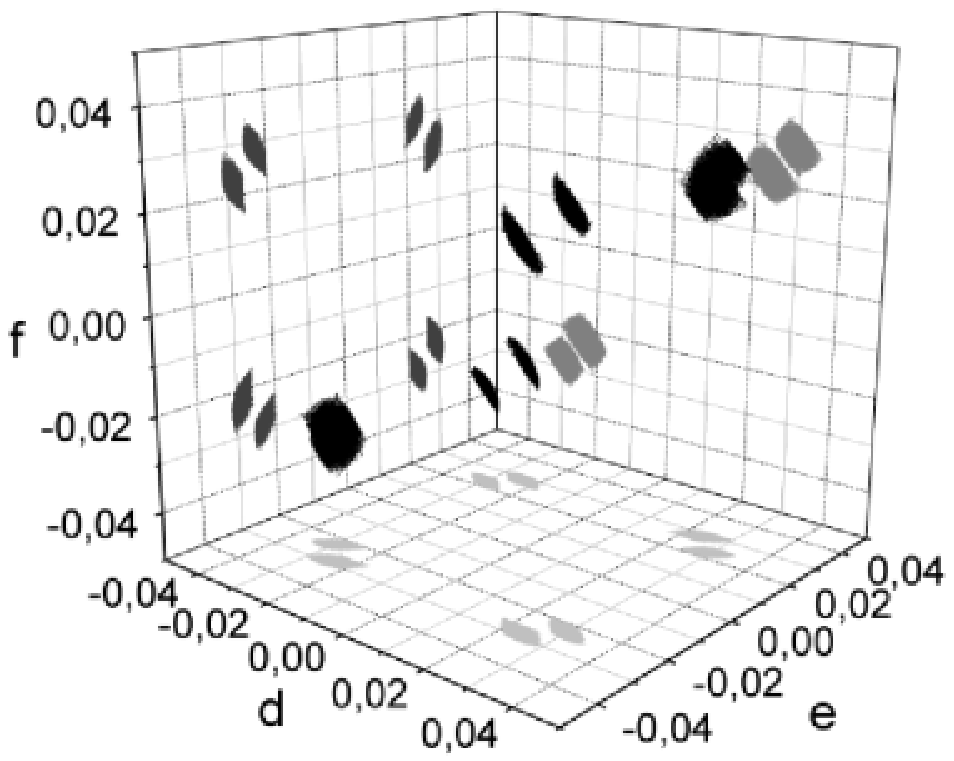}}
	\end{tabular}
	\caption{Allowed regions for the mass matrix with $a=0$ (A texture). The first row shows plots with $\alpha=1$ (present data, $3 \sigma$ level), 
the second row shows results for  $\alpha=2$.}
\label{azero}
\end{center}
\end{figure}

\begin{figure}[!ht]
\begin{center}
	\begin{tabular}{ccc}
		\resizebox{60mm}{!}{\includegraphics{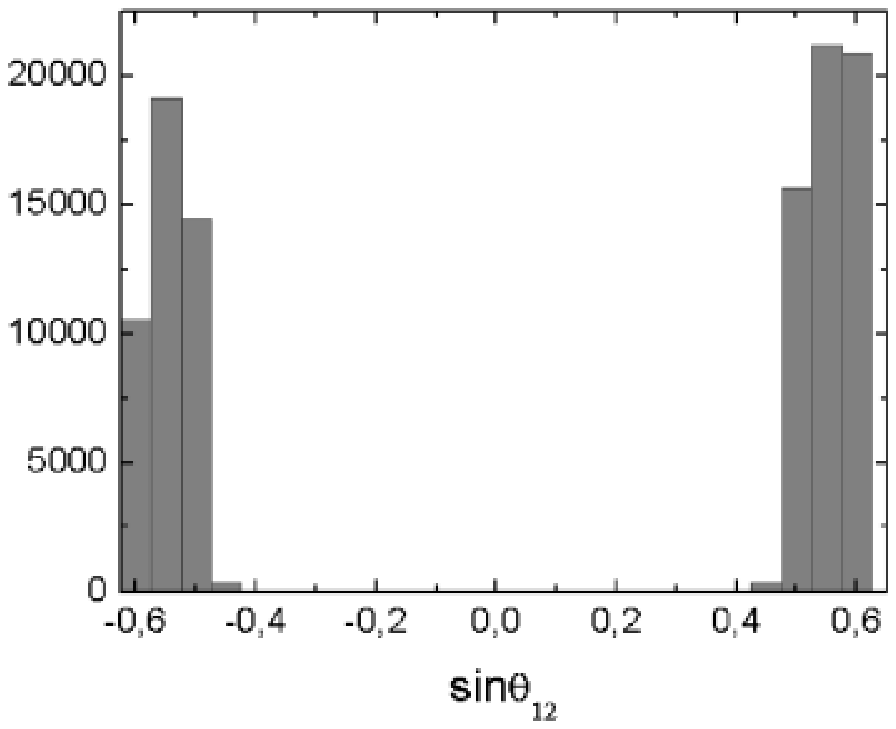}}&
		\resizebox{60mm}{!}{\includegraphics{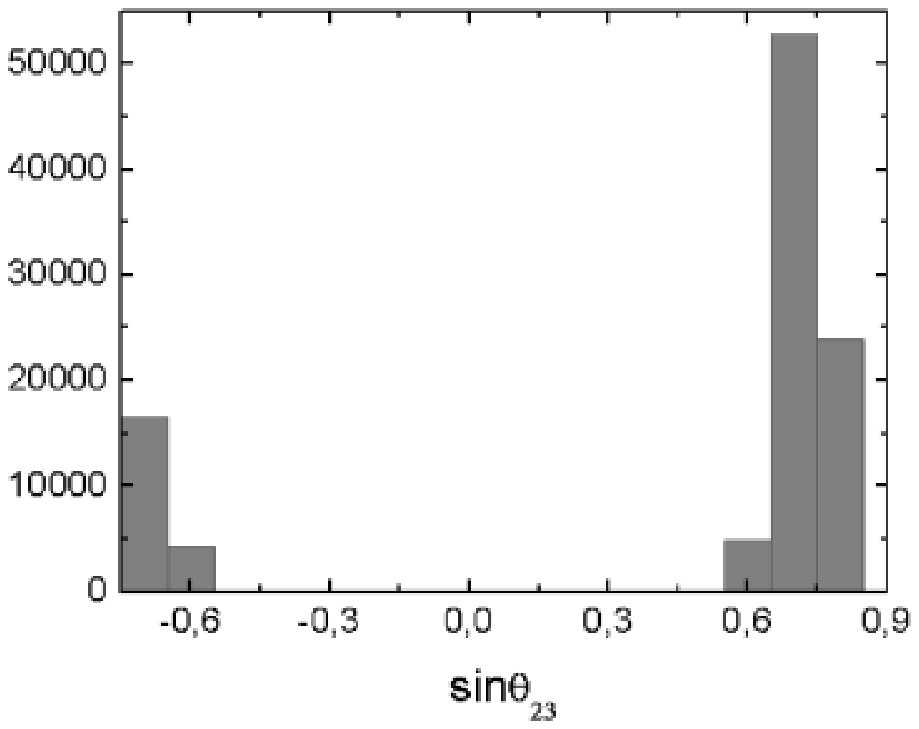}}&
		\resizebox{60mm}{!}{\includegraphics{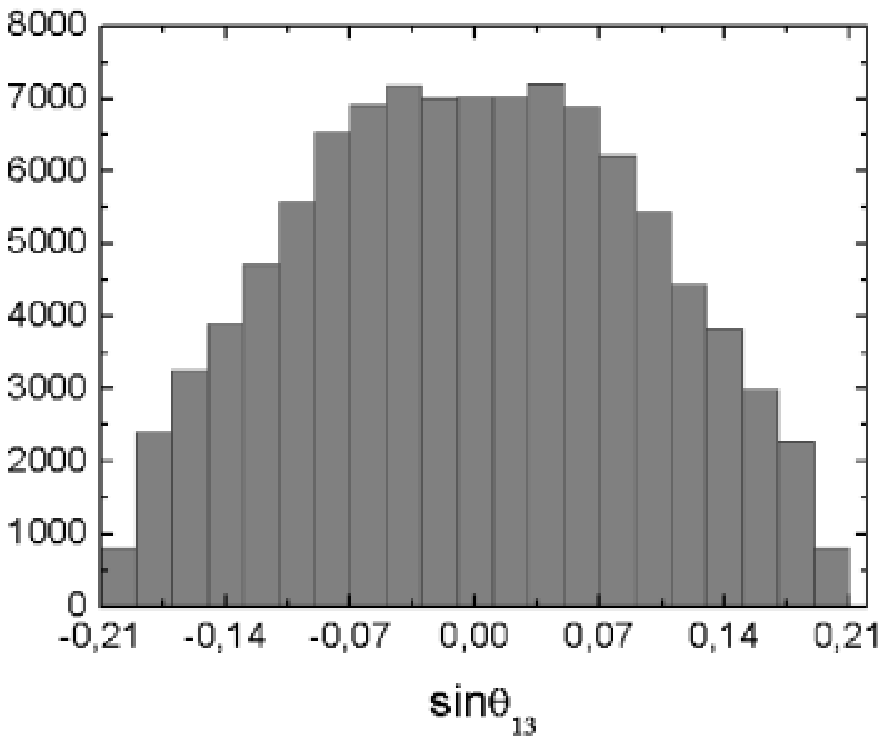}}
	\end{tabular}
	\caption{Histograms of neutrino rotation angles $\sin \theta_{12}$, $\sin \theta_{23}$ and $\sin \theta_{13}$ for neutrino mass texture with one zero $a=0$. The histogram for $\sin{\theta_{23}}$ does not depend on $a$ and
is the same for $a \neq 0$.}
\label{azeroH}
\end{center}
\end{figure}

\begin{figure}[!ht]
\begin{center}
	\begin{tabular}{ccc}
  	        \resizebox{80mm}{!}{\includegraphics{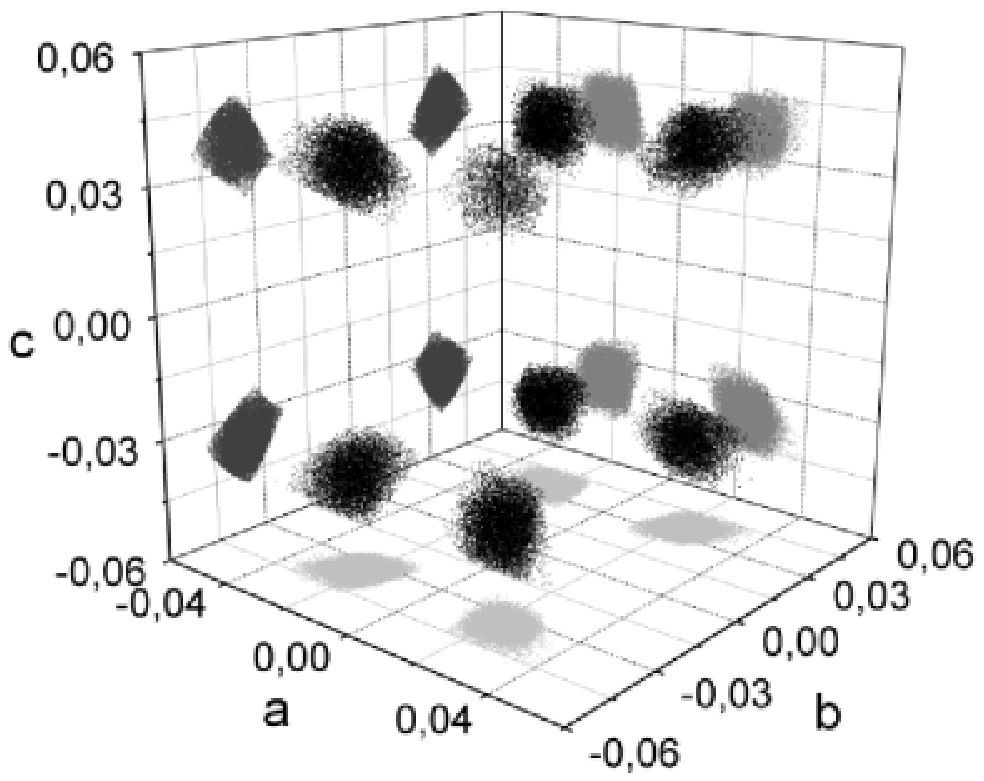}}&
		\resizebox{80mm}{!}{\includegraphics{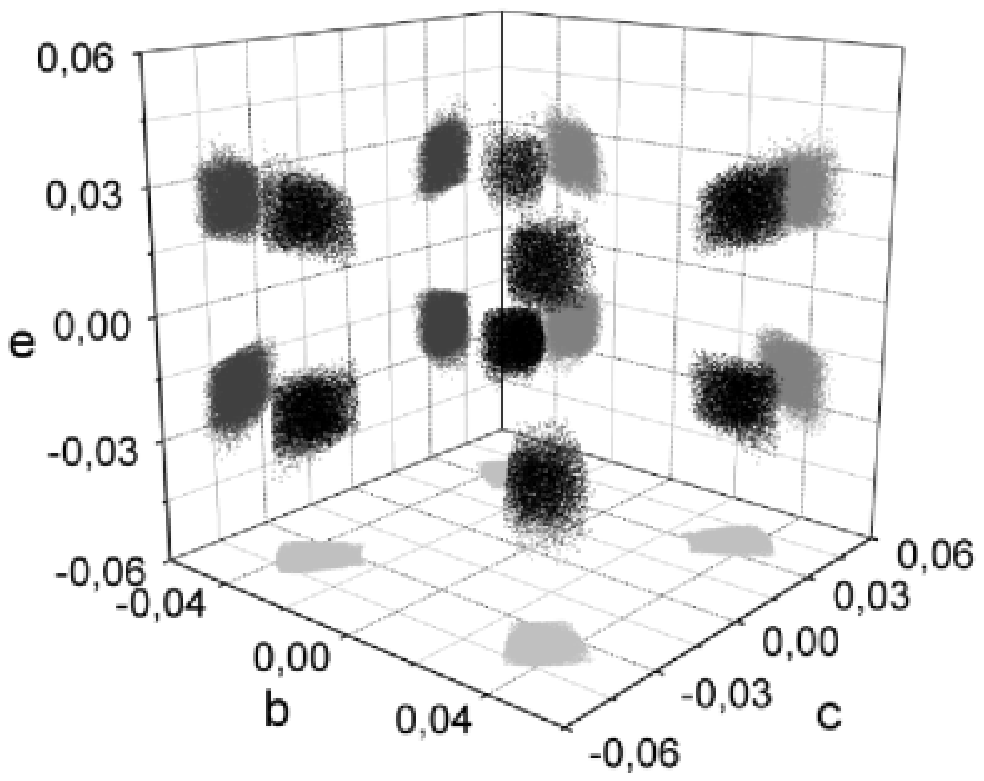}}\\
		\resizebox{80mm}{!}{\includegraphics{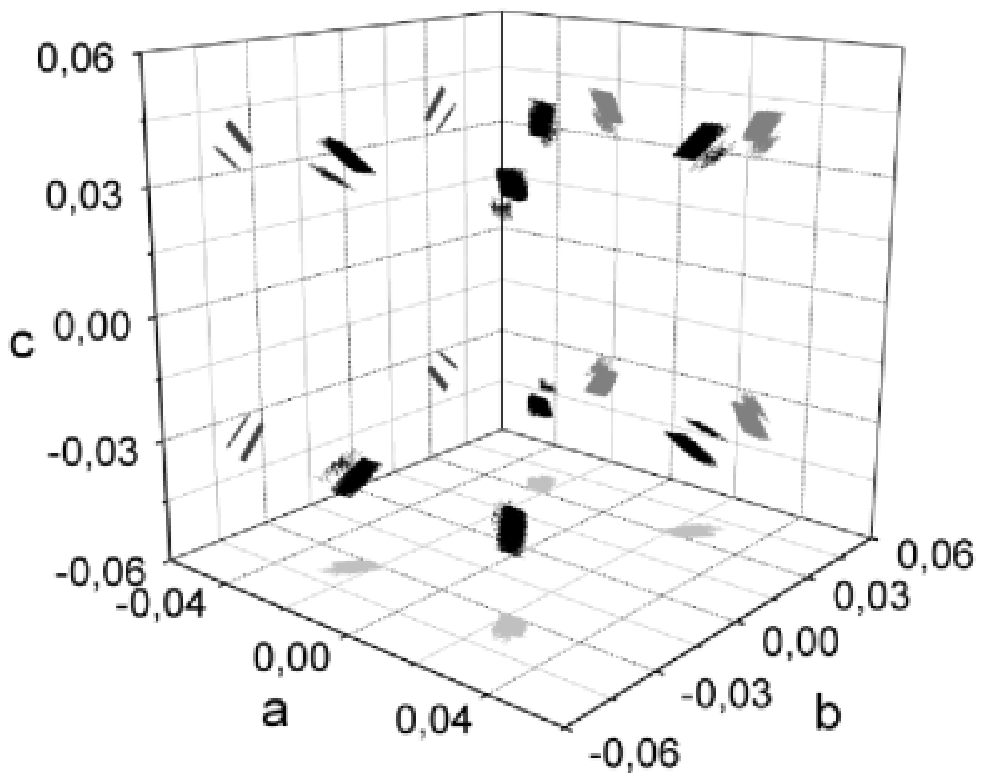}}&
		\resizebox{80mm}{!}{\includegraphics{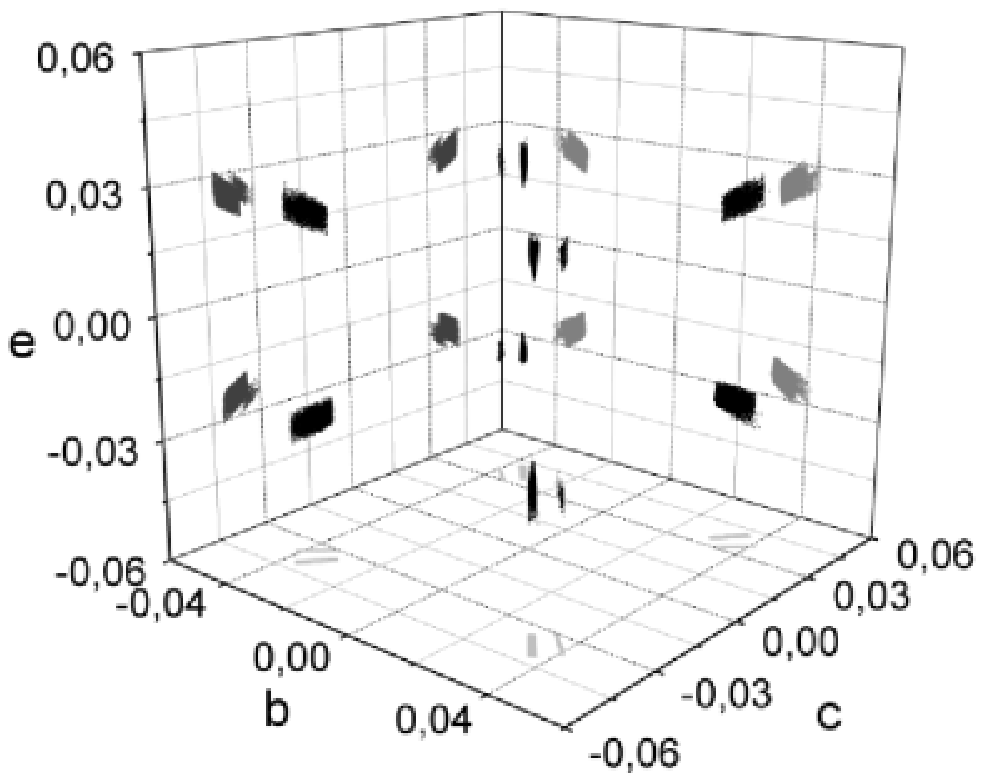}}
	\end{tabular}
	\caption{Allowed regions for the mass matrix with $d,f=0$ (C texture). The first row shows plots with $\alpha=1$ (present data, $3 \sigma$ level). The second row shows results for
$\alpha=2$.}
\label{Ctex}
\end{center}
\end{figure}

\begin{figure}[!ht]
\begin{center}
	\begin{tabular}{ccc}
		\resizebox{60mm}{!}{\includegraphics{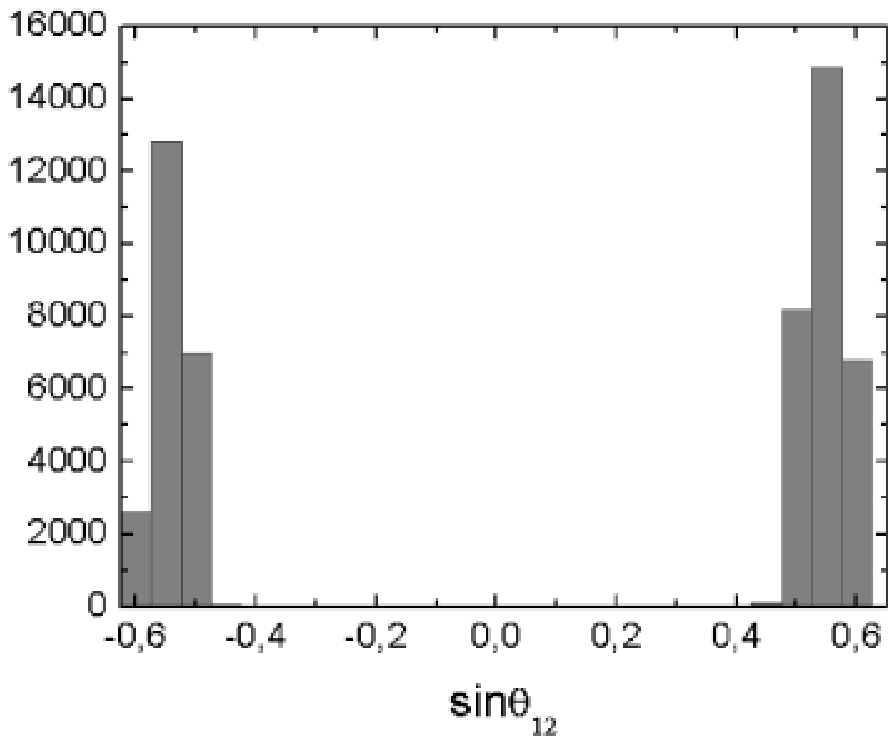}}&
		\resizebox{60mm}{!}{\includegraphics{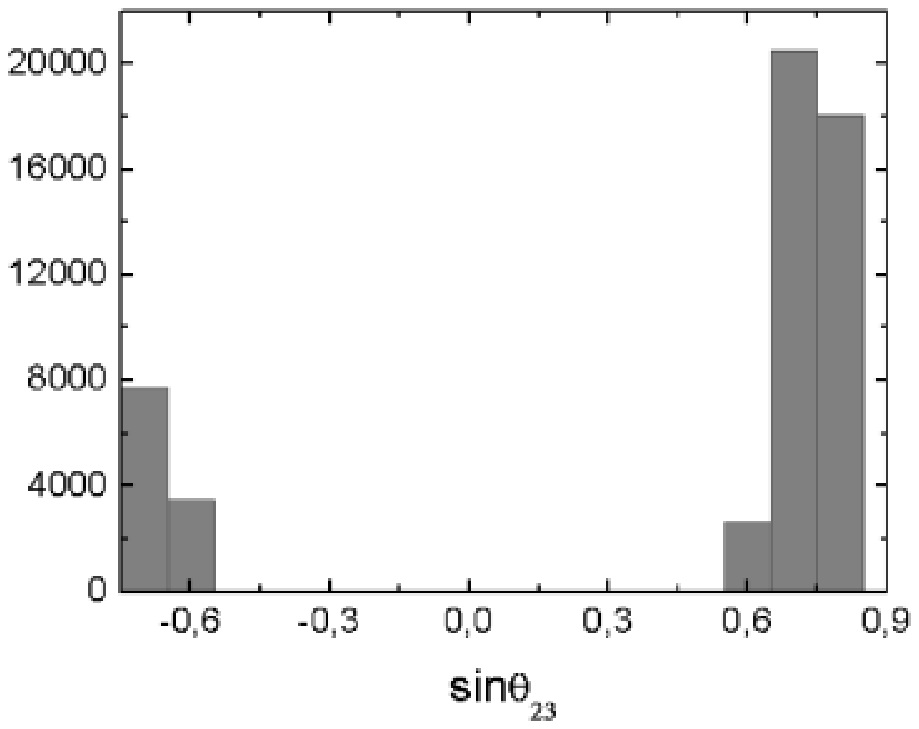}}&
		\resizebox{60mm}{!}{\includegraphics{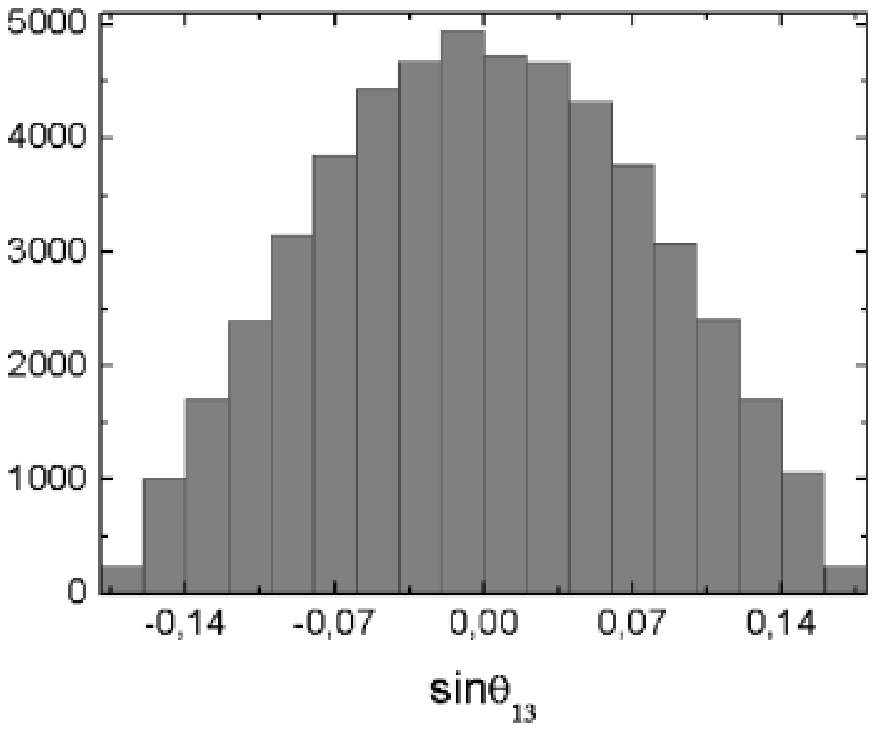}}
	\end{tabular}
	\caption{Histograms of neutrino rotation angles $\sin \theta_{12}$, $\sin \theta_{23}$ and $\sin \theta_{13}$ for neutrino mass texture with $d,f=0$ (texture C).}
\label{CtexH}
\end{center}
\end{figure}

\begin{figure}[!ht]
\begin{center}
	\begin{tabular}{ccc}
		\resizebox{60mm}{!}{\includegraphics{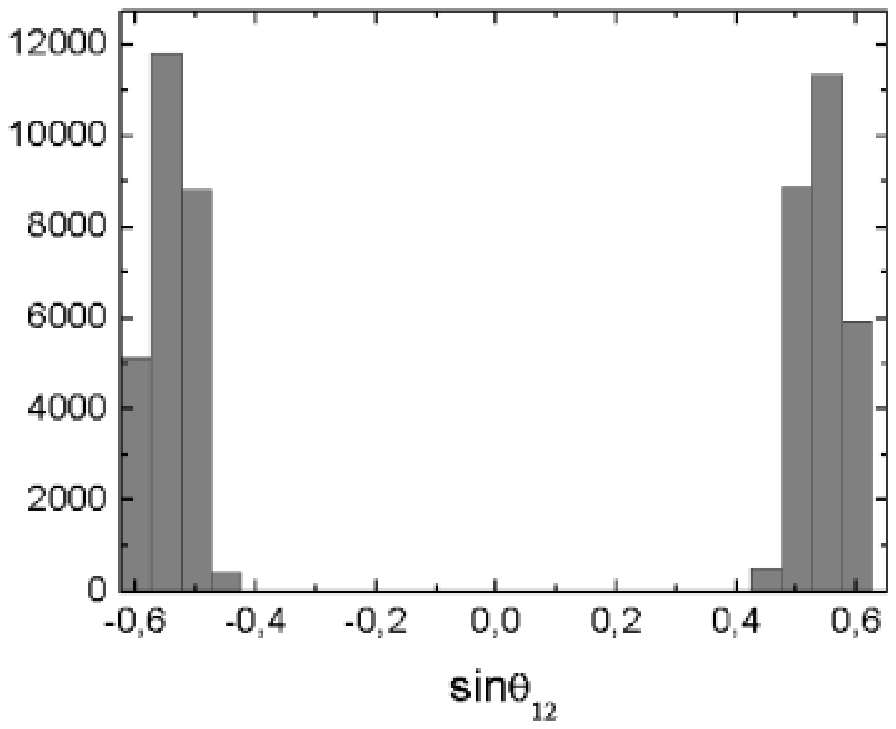}}&
		\resizebox{60mm}{!}{\includegraphics{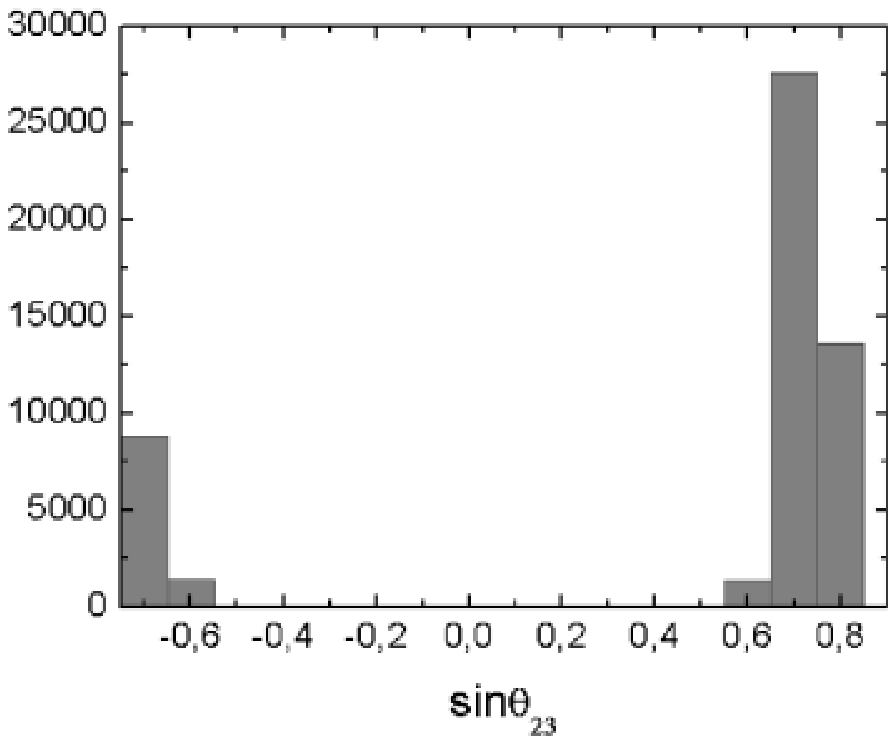}}&
		\resizebox{60mm}{!}{\includegraphics{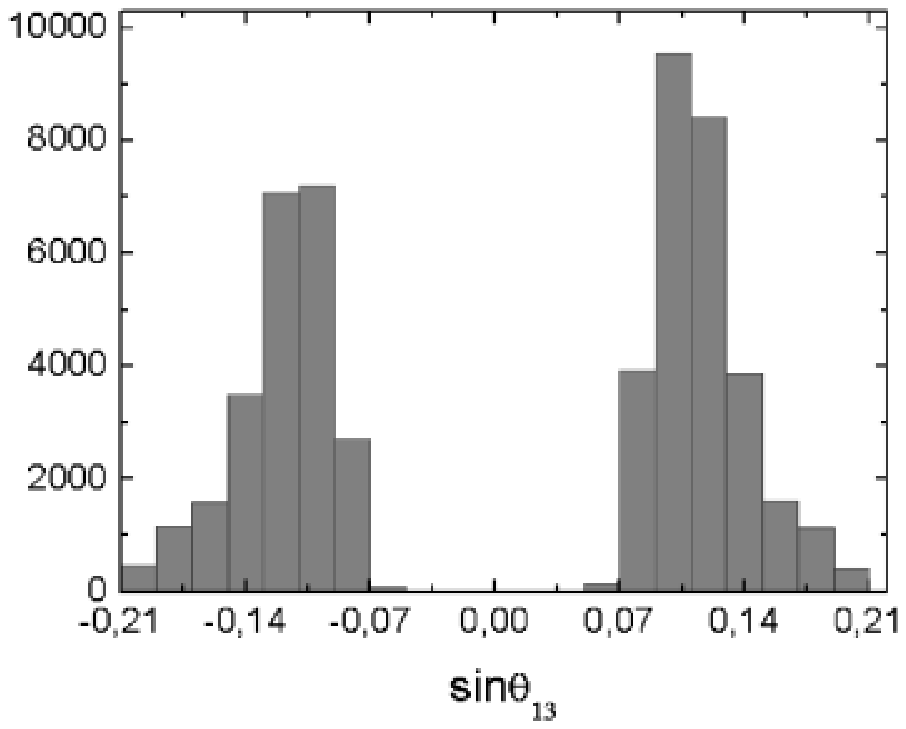}}
	\end{tabular}
	\caption{Histograms of neutrino rotation angles $\sin \theta_{12}$, $\sin \theta_{23}$ and $\sin \theta_{13}$ for neutrino mass texture with $a,b=0$ and $a,c=0$.}
\label{AH}
\end{center}
\end{figure}

\begin{figure}[!ht]
\begin{center}
	\begin{tabular}{ccc}
		\resizebox{60mm}{!}{\includegraphics{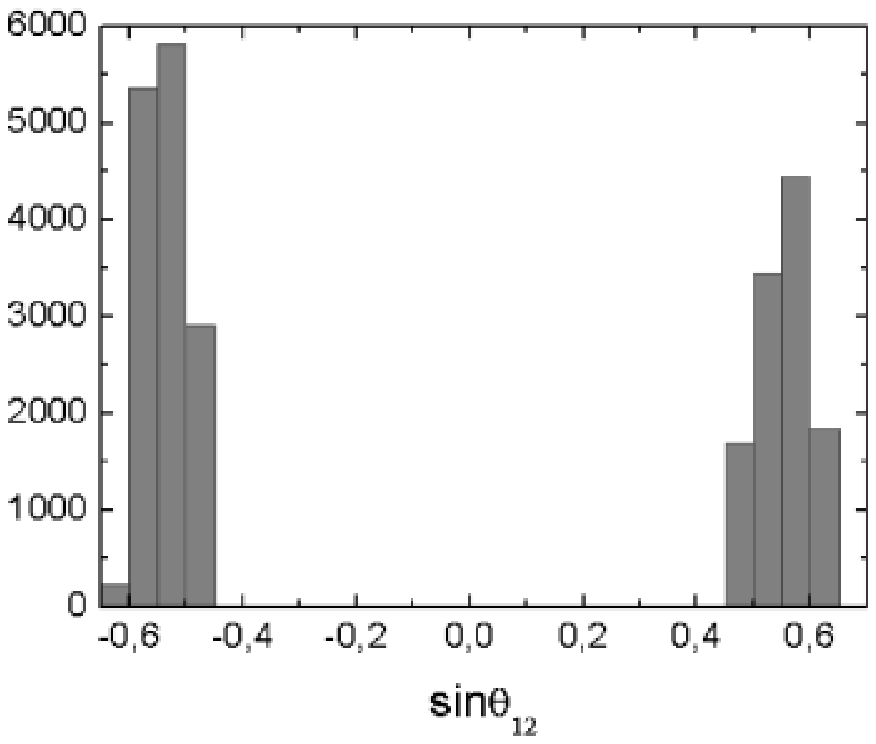}}&
		\resizebox{60mm}{!}{\includegraphics{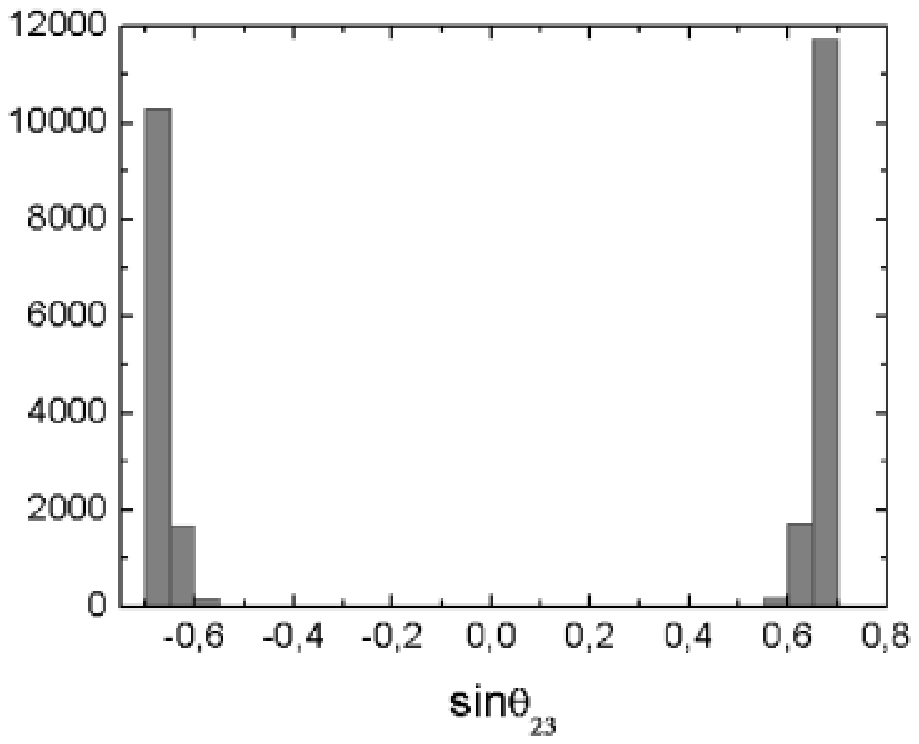}}&
		\resizebox{60mm}{!}{\includegraphics{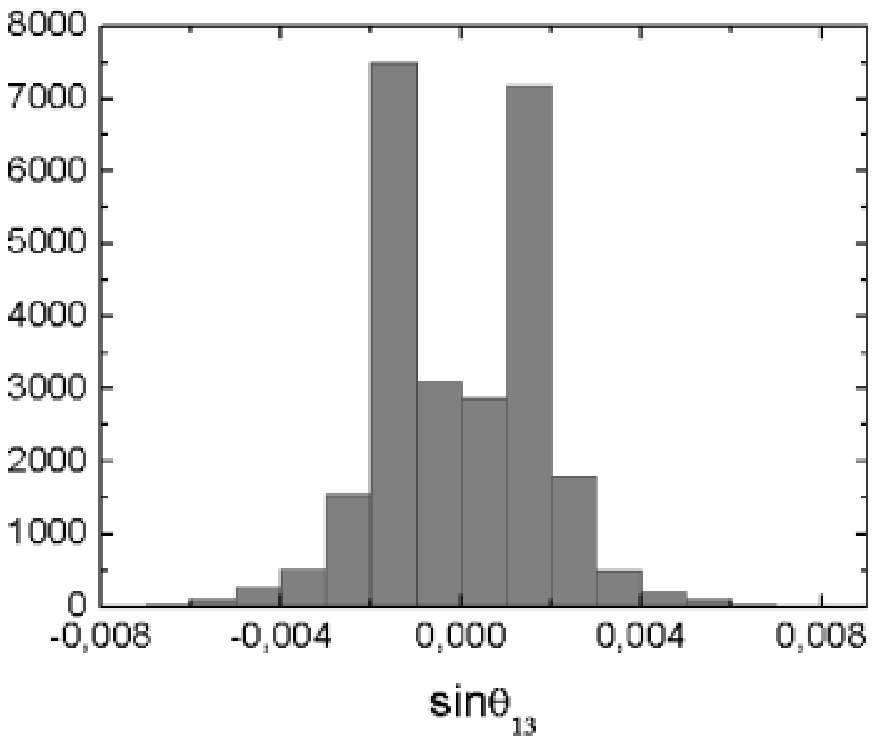}}
	\end{tabular}
	\caption{Histograms of neutrino rotation angles $\sin \theta_{12}$, $\sin \theta_{23}$ and $\sin \theta_{13}$ for neutrino mass textures $B$ (all cases).}
\label{BH}
\end{center}
\end{figure}

The main purpose of the algorithm is to find values of parameters of neutrino mass matrix $M_{\nu}$, Eq.~\ref{eq:generalcase}, which lead to results which are within ranges of current neutrino experimental data. It is obvious that it is not possible to check numerically the all possible combinations of input parameters, the Adaptive Monte Carlo (AMC) is the best solution. The program works in two steps. Firstly, it generates randomly input parameters and verifies the obtained output with experimental data. Thus the basic set of parameters is obtained. Such sets are represented as points on plots, see e.g. the left plot in Fig.~\ref{mc}. 
The user can specify how many points he wants to obtain. For a typical run it is $N_{Scat}=100$. As soon as the first step ends,  AMC algorithm begins. It uses points which are obtained during the first step and searches for a solution with 
the lowest value of  $\chi^{2}$ around every of these points. Second step consists of $N_{it}$ iterations (enlargements), 
and for every iteration a diagonalization of the neutrino mass matrix is performed by generating randomly points $N_{calc}$ times. 
As a result of the second step, additional points are obtained which form regions of allowed parameter space, e.g. the right plot in Fig.~\ref{mc}. One can notice that these regions look denser (more points) in the center than near the edges. It is a result of an algorithm which zooms into the points with the lowest values of $\chi^{2}$. Below is a sketch of the algorithm:\\

%
%
\begin{enumerate}
\item First step: Scattering
       \begin{itemize}
      	 \item \emph{Random generation of input parameters}\\
For the nth iteration $(n=1,2,\ldots)$ the program generates set of input parameters: 
neutrino mass matrix elements. Algorithm stops the loop when $n=N_{Scat}$. Because in this work we consider Majorana neutrinos then neutrino mass matrix is symmetric so this 
implies that diagonal elements of the mass matrix are real.
       	 \item \emph{Diagonalization of the neutrino mass matrix $M_{\nu}$}\\ 
With mass matrix input parameters chosen, the algorithm finds eigenvectors (neutrino masses) and normalized eigenstates 
(columns of the neutrino mixing matrix) 
       	  \item \emph{Comparison with experimental results and saving allowed parameters}\\
The program compares the obtained values with experimental results (neutrino mass squared differences and absolute values 
of neutrino mixing matrix elements) with a help of the $\chi_i$ function  
\begin{equation}
	\chi_{i}^{2}=\frac{(x_{i}^{cent}-x_{i})^2}{ \big( \frac{\sigma_{i}}{\alpha} \big) ^{2}},
\label{eq:A1}
\end{equation} 
where $x_{i}$ is a calculated value and $x_{i}^{cent}$, $\sigma_{i}^{2}$ are experimental best fit and uncertainty values, 
respectively. 
The constant $\alpha$ is a parameter which  controls decreasing
of an assumed error, Eq.~\ref{eq:ch}. 
If calculated $\chi_i^2$ is smaller than the demanding number, the parameters (points) are saved.
         \end{itemize}
The purpose of this step is to scatter randomly points defined by the neutrino mass matrix parameters in a hyperspace volume 
where a number of dimensions is equal to  the number of input parameters. 
%
%
\item Second step: The Adaptive Monte Carlo
       \begin{itemize}
       \item \emph{Reading obtained points}\\
For the $n$-th iteration $(n=1,2,\ldots ,N_{Scat})$ program reads set of  parameters (points) obtained in the previous step 
and sets their values as the central values $x_{i}^{cent}$. 
For the $i$-th iteration $(it=1,2,\ldots ,N_{it})$, the  algorithm sets range of input parameters 
(which will be randomly generated in the next step):
\begin{equation}
	x_{i}^{cent}\pm \xi_{it} \delta_{i},  
\label{eq:A2}
\end{equation}
where $i$ numerates input parameters, $\delta_{i}$ is an initial range and $\xi_{it}$ is a function defined as follows:
\begin{equation}
	\xi_{it}=\left\{
	\begin{array}{ll}
	1 & it=0,\\
	0.6/it & it>0.
	\end{array} \right.
\end{equation}
The constant number  $0.6$ and the function $1/it$ are set empirically.
%
%
Similarly as in the first step, the program executes the following (a)-(c) tasks $N_{calc}$ number of times:
		\begin{itemize}
				\item[(a)] \emph{Random generation of input parameters}
				\item[(b)] \emph{Diagonalization}
				\item[(c)] \emph{Comparison with experimental data and saving successive cases}
			\end{itemize}
		      \item \emph{Setting new central values}\\
When tasks $a)-c)$ are finished, the algorithm chooses these set of input parameters for which the calculated value of $\chi^{2}$ 
is the lowest:
\begin{equation}
	\chi^{2}=\sum_{i=1}^{11}\chi_{i}^{2}.
\label{eq:A4}
\end{equation}
and sets it as  "the best set", i.e. as a new central value $x_{i}^{cent}$. Next, the program repeats above calculations with the new 
value of $it$ which is $it+1$. Note that while increasing $it$, the value of function $\xi_{it}$ decreases, 
so the range in which we randomly generate input parameters will be smaller and smaller as $it$ increases. 
Also, since every time the best $\chi^{2}$ is chosen,  the solution moves towards the most probable solution 
for a considered set of experimental results.  
		\end{itemize}
\end{enumerate}






\end{document}